\documentclass[aip,graphicx,floatfix]{revtex4-1}
\usepackage{graphicx}

\draft 

\begin{document}

\title{60 Gbps real-time wireless communications at 300 GHz carrier using a Kerr microcomb}

\author{Brendan M. Heffernan}
    \affiliation{IMRA America Inc., Boulder Research Labs, 1551 South Sunset St, Suite C, Longmont, Colorado 80501, USA}
\author{Yuma Kawamoto}
\author{Keisuke Maekawa}
    \affiliation{Information Photonics Group, Div. Adv. Electronics and Optical Science, D348 Graduate School of Engineering Science, Osaka University, 1-3 Machikaneyama, Toyonaka, 560-0043 Osaka, Japan}
\author{James Greenberg}
\author{Rubab Amin}
    \affiliation{IMRA America Inc., Boulder Research Labs, 1551 South Sunset St, Suite C, Longmont, Colorado 80501, USA}
\author{Takashi Hori}
\author{Tatsuya Tanigawa}
    \affiliation{IMRA America Inc., Japan Branch Office, 2-1 Asahi-machi, Kariya, Aichi, Japan}
\author{Tadao Nagatsuma}
    \affiliation{Information Photonics Group, Div. Adv. Electronics and Optical Science, D348 Graduate School of Engineering Science, Osaka University, 1-3 Machikaneyama, Toyonaka, 560-0043 Osaka, Japan}
\author{Antoine Rolland}
    \affiliation{IMRA America Inc., Boulder Research Labs, 1551 South Sunset St, Suite C, Longmont, Colorado 80501, USA}
    \email[]{arolland@imra.com} 

\date{\today}

\begin{abstract}
Future wireless communication infrastructure will rely on terahertz systems that can support an increasing demand for large-bandwidth, ultra-fast wireless data transfer. In order to satisfy this demand, compact, low-power, and low noise sources of terahertz radiation are being developed. A promising route to achieving this goal is combining photonic-integrated optical frequency combs with fast photodiodes for difference frequency generation in the THz. Here, we demonstrate wireless communications using a 300 GHz carrier wave generated via photomixing of two optical tones originating from diode lasers that are injection locked to a dissipative Kerr soliton frequency microcomb. We achieve transfer rates of 80 Gbps using homodyne detection and 60 Gbps transmitting simultaneously both data and clock signals in a dual-path wireless link. This experimental demonstration paves a path towards low-noise and integrated photonic millimeter-wave transceivers for future wireless communication systems.
\end{abstract}

\maketitle 

\section{Introduction}
Modern society's insistence on faster wireless data transfer necessitates increased bandwidth. However, overcrowding at currently deployed channels means that novel carrier frequencies must be developed to meet demand. The teraherz domain (which we take to primarily mean 100 GHz - 1 THz in the context of wireless communications) is largely untapped. With plentiful bandwidth for high-speed communications, this spectral region could provide a solution to current congestion.~\cite{Kurner2022} In fact, this regime has already been targeted for the proposed sixth generation of wireless communication standards (6G). \cite{Dang2020} This target is enabled by recent advances in the creation and detection of THz radiation using electronic, photonic, and hybrid approaches.~\cite{Sengupta2018} While all approaches have merit and are progressing quickly, hybrid (or photonics assisted) sources have already shown considerable promise for communications.~\cite{Pang2022, Yao2022, Jia2022}

In particular, photomixing two laser lines separated by a THz frequency on a suitably fast photodiode provides several key benefits. Generally, this method is readily integrated with existing optical fiber technology and provides the widest frequency tunability and finest frequency resolution available of any THz generation technology.~\cite{Safian2019} Writing a data stream on an optical signal can be realized in a compact and low-cost fashion using an electro-optic modulator up to 100~GHz. For very large carrier frequencies, for instance up to 300~GHz, plasmonic modulators may be used.~\cite{Ummethala2019} Another benefit is the level of frequency stability that can be achieved. Optical frequencies can be stabilized to unmatched levels, and photomixing transfers this low noise to the THz regime. Unrivaled fractional frequency stabilities of $10^{-16}$ at one second have been reached in the microwave domain up to 100~GHz,~\cite{Fortier2016, Xie2017, Nakamura2020} and similar techniques have recently been applied to the THz region to produce record-low phase noise. \cite{Tetsumoto2021} Low phase noise terahertz sources generated from light with high optical signal-to-noise ratios (OSNR) are important in communications for implementing dense modulation schemes, such as high order quadrature amplitude modulation (QAM). \cite{Chen2018} Here we use the term OSNR to refer to the optical light before modulation in order to differentiate it from the signal-to-noise ratio of the THz wave at the receiver.

Generating spectrally pure THz radiation through photomixing can be accomplished in a straightforward manner using two low-noise diode lasers.\cite{Pang2011, Nagatsuma2013, Harter2020} However, these lasers are quite expensive, and they are not phase and frequency stabilized with respect to one another, leading to added noise in the resulting THz waves and difficulty in implementing heterodyne detection schemes. Therefore, the benefits of using optical frequency combs (OFC's) in this application have long been recognized. \cite{Nagatsuma2013} Two lines from an OFC can be filtered and used to encode data while their shared source mitigates environmental noise at the transmitter. Previously, OFC's produced via mode-locked lasers and electro-optic modulators have been used to successfully demonstrate large wireless data transmission rates at various THz carriers. \cite{Kanno2011, Nagatsuma2013, Koenig2013, Mohammad2018a} Depending on the generation method, OFC's can suffer from limited power and OSNR in a given comb tooth. This, in turn, degrades the spectral purity of the THz. Conventional erbium-doped fiber amplifiers (EDFA's) combined with optical filtering may alleviate this problem in some situations, but this involves multiple steps and, until very recently, was not miniaturizable. \cite{Liu2022} An alternative approach for amplifying comb modes is provided by optical injection locking (OIL). \cite{Liu2020a}

OIL is equivalent to injection locking in classical oscillators and has been investigated since the invention of the laser. \cite{Stover1966} Light from one oscillator is incident into the cavity of a second, causing stimulated emission that is locked to the phase of the first. Under the correct conditions, this enables low power single frequency sources to be amplified by up to 70 dB with little added phase noise. \cite{Kakarla2018} Furthermore, with proper optimization of injected power, the linewidth of the high-power laser will converge to the injected linewidth, making it possible to use cheap, multimode diodes for amplification. \cite{Liu2020a} OIL has been studied previously using optical frequency combs generated from mode locked lasers, \cite{Cai1995, Moon2006, Ryu2010} cascaded electro-optic modulators, \cite{Balakier2014, Shortiss2018} and microresonators, \cite{Kuse2022} as well as from a theoretical perspective, \cite{Gavrielides2014, Doumbia2020} establishing the dynamics of locking, amplification and noise transfer. The simultaneous filtering and amplification properties of OIL have made injection locking laser diodes (LD's) to OFC's increasingly popular for THz communications experiments in recent years. It was demonstrated that an OIL-based approach was more suited to communications than simple filter and amplify configurations in low optical signal to noise ratio situations.~\cite{Albores-Mejia2015} 

Many aspects of THz transmitters based on LD's injection locked to OFC's are imminently miniaturizable on photonic integrated circuit (PIC's). The LD's and uni-carrier traveling photodiodes (UTC-PD's) \cite{Ishibashi1997, Ishibashi2020} on which the optical tones are mixed have been integrated, along with important modulation components. \cite{Balakier2014} Similarly, laser integration has been demonstrated in offline communications experiments reaching 130 Gbps with a 300 GHz carrier. \cite{Jia2022} Creating compact and energy efficient THz transmitters will be key for future commercialization and deployment. However, to date, these demonstrations have relied on bulky, external comb formation.

Dissipative Kerr soliton (DKS) frequency combs can be formed through non-linear optical processes in micron-scale ring resonators. \cite{Kippenberg2018} In a material such as silicon nitride, they can have comb mode spacing from tens of gigahertz to a terahertz, are compatible with wafer-scale fabrication, can be compact and low power,\cite{Shen2020} and can be packaged with LD's, modulators and UTC's via photonic wire bonding\cite{Billah2018} or micro-transfer-printing, \cite{Maes2023} making them a strong candidate for low noise THz sources. In fact, a very recent proof of concept experiment has shown that a Kerr comb-based source can support complex modulation formats up to 64 QAM. \cite{Tetsumoto2022} However, this demonstration relied on careful dispersion compensation to coherently modulate 10 comb teeth at once in order to overcome limited OSNR on each individual tooth and the baud rate was very limited.

\begin{figure}[ht!]
\centering\includegraphics[width=5in]{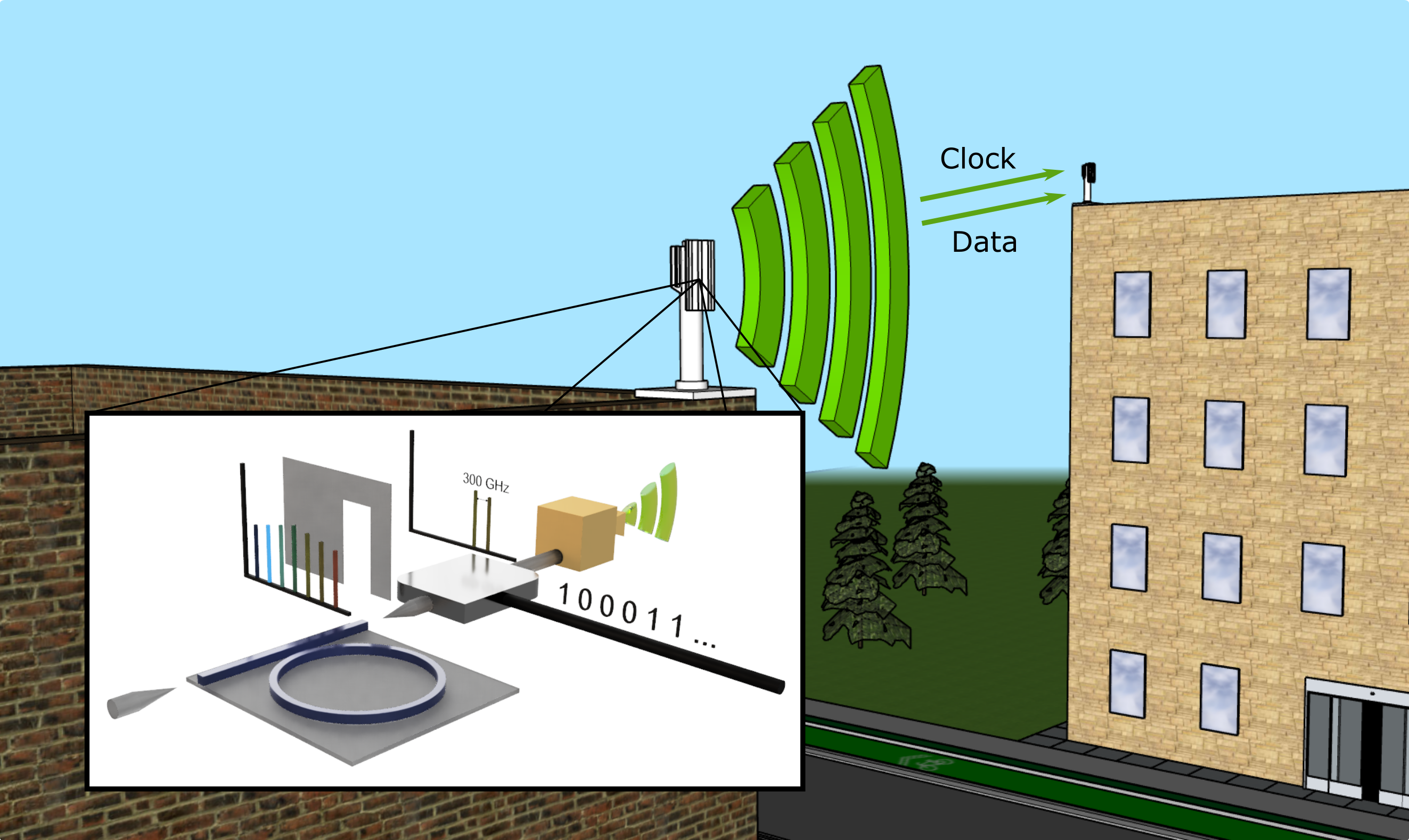}
\caption{\label{concept}  A conceptual figure demonstrating high speed wireless communication with a 300 GHz carrier enabled by a microcomb.}
\end{figure}

Here we introduce and characterize a source based on LD's injection locked to a Kerr microcomb. Counter to prior work, \cite{Tetsumoto2022} this source does not require dispersion compensation, simplifying the architecture for broader adoption. We then use this source to demonstrate wireless communication at 300 GHz and a configuration-dependent transfer rate of 60 or 80 gigabits per second (Gbps). We achieve 60 Gbps using real-time processing in a practical communication scheme where the transmitter and receiver are completely separated; both the data stream and local oscillator are wirelessly transferred, as pictured conceptually in Fig. \ref{concept}.

\section{Millimeter Wave Source}

\begin{figure}[ht!]
\centering\includegraphics[width=6.0in]{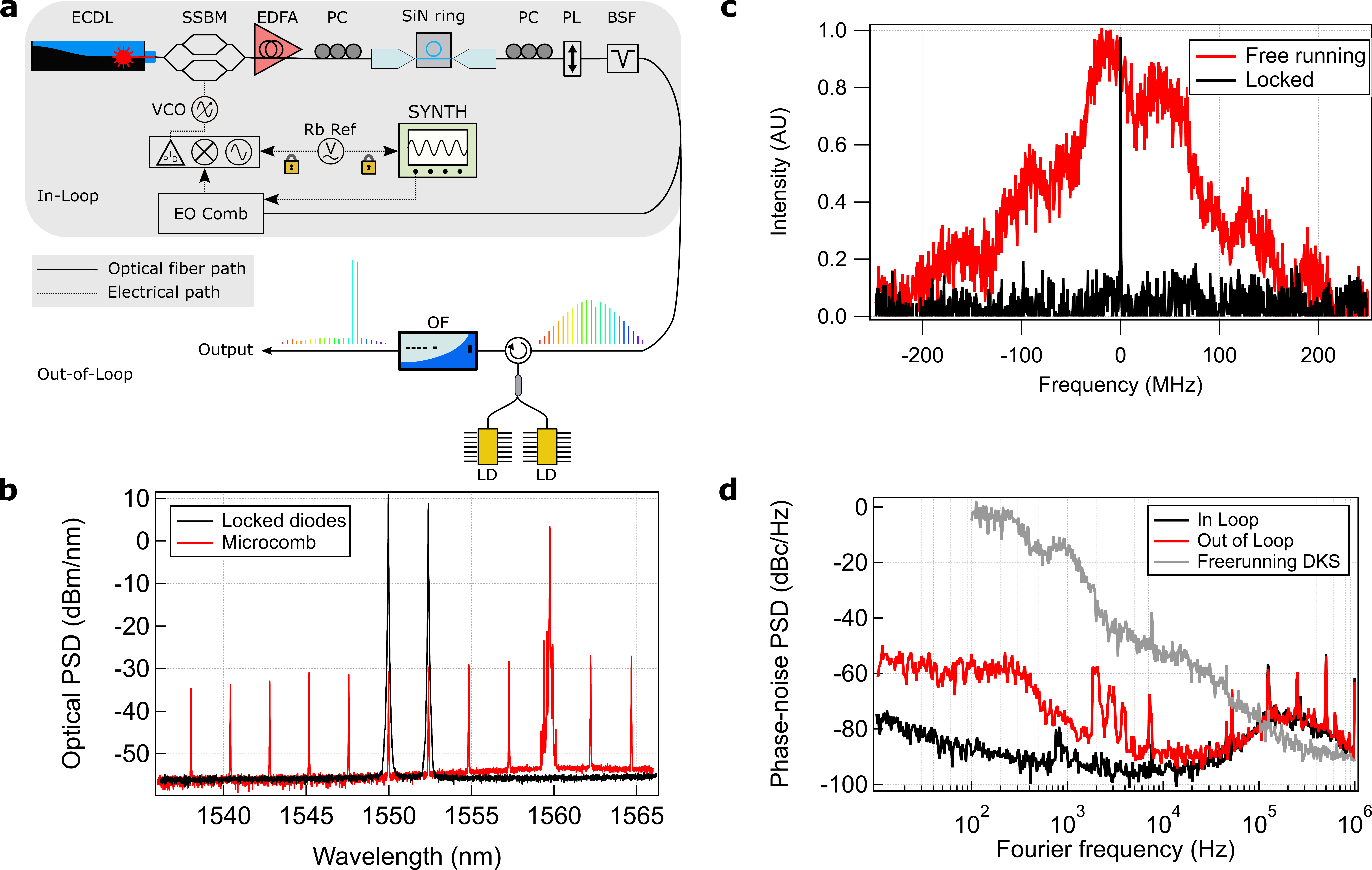}
\caption{\label{microwave}  \textbf{a}, Schematic of the microcomb-based source to generate THz radiation via photomixing. ECDL, external cavity diode laser; SSBM, single sideband modulator; EDFA, erbium-doped fiber amplifier; PC, polarization controller; PL, polarizer; SiN ring, silicon nitride microresonator; BSF, band stop filter; LD, laser diode; VCO, voltage controlled oscillator; Rb Ref, rubidium clock reference; SYNTH, frequency synthesizer; EO comb, electro-optic comb; OF, optical filter. Light from the ECDL was frequency shifted by the SSBM before amplification. The light was then coupled to the SiN resonator to generate a Kerr comb. The pump light was filtered from the comb by a BSF before being split along two paths. The first (upper) path was used to read out and feed back on the repetition rate of the microcomb. The second (lower) branch was used to inject two laser diodes and filter the resulting output. \textbf{b}, Optical spectra of the microcomb (red) and injection locked LD's after filtering (black). \textbf{c}, RF beatnote between the two optical tones of the LD's as measured using an EO free-running (red) and injection locked to a comb teeth (black). The data was acquired using a 30 kHz bandwidth and averaged over 20 traces. A massive reduction in linewidth is clearly evident. \textbf{d}, Phase noise of the in loop DKS comb (black) and the out of loop injection-locked diodes (red), locked to a reference oscillator compared to the phase noise of the free-running DKS comb repetition rate (grey).}
\end{figure}

The microcomb-based THz source is shown in Fig.\ref{microwave} \textbf{a}. An external cavity diode laser (ECDL) (ECDL CTL1550, Toptica Photonics) pumped the microcomb. The pump was initially sent to a carrier-suppressed single side band modulator (CS-SSBM) that can sweep the frequency at a rate of roughly 100 GHz/µs over a 2 GHz span to reliably initiate the microcomb in a single-soliton regime. \cite{Stone2018} The pump was amplified and edge coupled into an on-chip bus waveguide using a lensed fiber. The DKS comb was generated in a silicon nitride ring and a second lensed fiber acted as the output coupler. After the DKS comb was generated, the output was sent through a band stop filter (BSF) to remove the residual pump light. The DKS comb was then split into two paths. In the OIL path, only one circulator and no optical band pass filters were used to minimize the non-common-path of the split comb light. A 50:50 splitter/coupler was used on port two of the circulator. The entire comb was injected into both LD's without detriment because the spacing of the injected comb teeth was much greater than the locking bandwidth. \cite{Cai1995} Approximately -30 dBm of power in the relevant comb line was injected, leading to an injection ratio of -40 dB. The side mode oscillations of the LD's were then filtered using a waveshaper. The resulting optical spectrum of the injection-locked LD's compared to the DKS is shown in Fig. \ref{microwave} \textbf{b}. The injection-locked LD's show a 35 dB improvement in OSNR compared to the microcomb as measured with a .02 nm resolution. 

We tested the noise transfer of the microcomb repetition rate to the frequency difference of the LD's by sending a copy of the microcomb output directly into an electro-optic (EO) comb consisting of cascaded electro-optic modulators (EOM's). The EOM's were driven at 10.0335 GHz and created a number of sidebands that spanned the $\approx$ 301 GHz gap between laser lines. This provided a photodetected signal in the hundreds of megahertz range that was used to read out the frequency difference between adjacent comb teeth. \cite{Metcalf2013} We compared the linewidth of this RF beatnote with and without injection locking, as shown in Fig. \ref{microwave} \textbf{c}. In the absence of OIL, the frequency difference between the two LD's is unstable and drifts, whereas locking to a frequency comb greatly reduce the linewidth. To further reduce frequency drift and characterize the phase noise, the detected beat note from this EO comb was divided down and sent to a mixer, where it was mixed with the output of a frequency synthesizer. The synthesizer was locked to a rubidium clock standard. The output of the mixer was an error signal fed to a PID controller with a narrow bandwidth. It fed back to the DKS comb pump frequency through a voltage controlled oscillator (VCO) driving the CS-SSBM, loosely locking the repetition rate of the DKS microcomb to mitigate drift. The injection-locked laser lines could be sent to an EO comb of a similar design for readout. The readouts of the in-loop microcomb and the out-of-loop OIL LD's could be sent to a signal analyzer (PXA Signal Analyser N9030A, Agilent Technologies) for analysis. 

The resulting in-loop and out-of-loop phase noise is shown in Fig. \ref{microwave} \textbf{d}. The out-of-loop phase noise represents the residual noise, i.e. the noise added by the injection-locking process and the associated setup. It is around -57 dBc/Hz at 100 Hz and -89 dBc/Hz at 10 kHz, while the in-loop phase noise is -87 dBc/Hz at 100 Hz and -95 dBc/Hz at 10 kHz. The error level of the in-loop phase is limited primarily because of the feedback bandwidth of the PID and the OSNR of the optical beatnote used for detection. We attribute the increased noise of the out-of-loop PSD to non-common fiber and distinct EO combs used to measure the frequency difference of the laser lines. Regardless, the added phase noise is well below the phase noise of the free-running microcomb, indicating that the OIL process is not degrading the phase noise of the microcomb repetition rate.



\section{Wireless Communication Demonstration}

\begin{figure}[ht!]
\centering\includegraphics[width=3.3in]{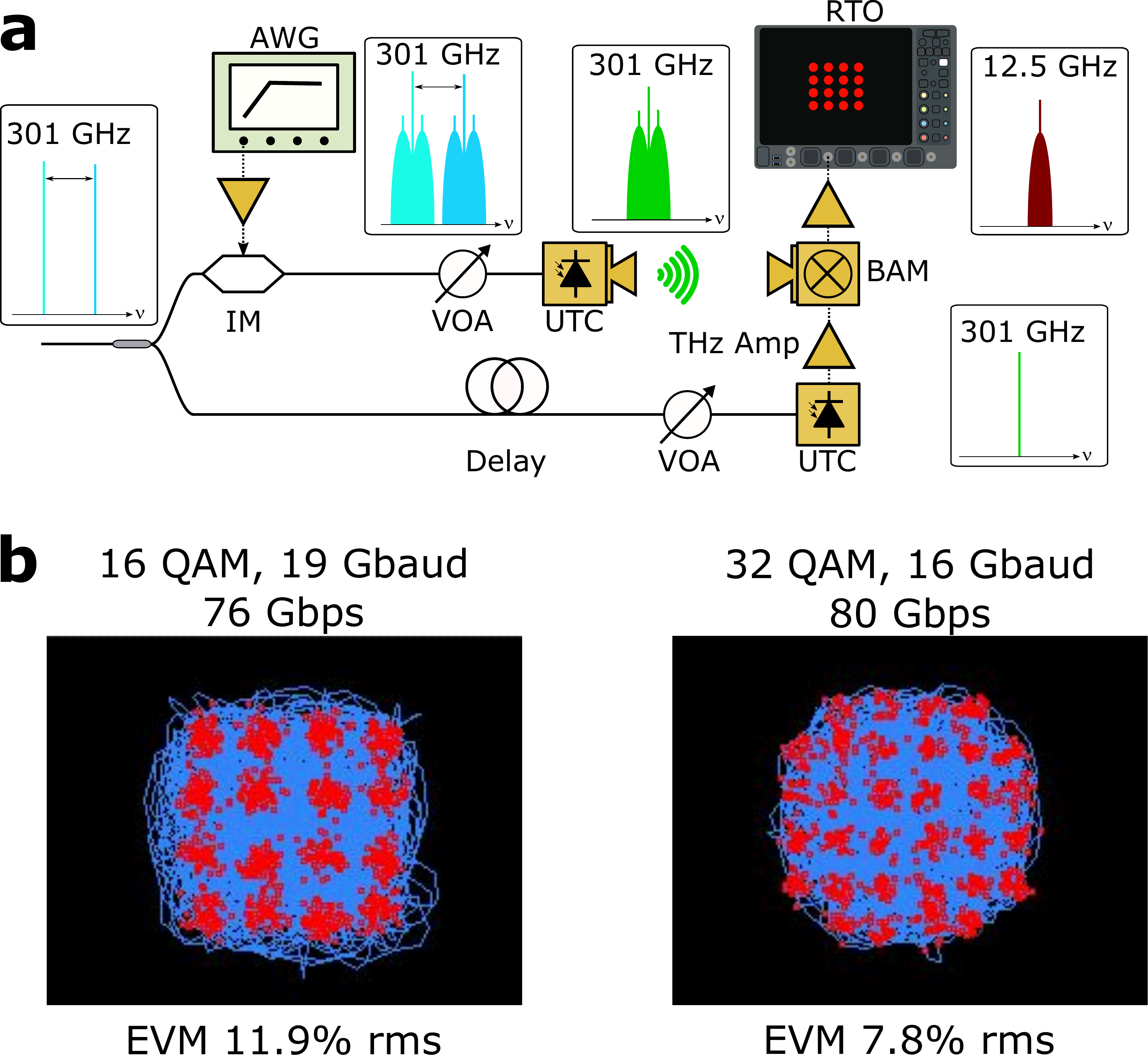}
\caption{\label{80gbps} \textbf{a}, Schematic layout of the homodyne wireless communication demonstration. Two laser lines from the source depicted in Fig. \ref{microwave} \textbf{a} are split. Approximately 80\% of the light is incident on an intensity modulator (IM) driven by an arbitrary waveform generator (AWG). The modulated light is attenuated before impinging on a UTC-PD. The resulting THz radiation is mixed down on a Schottky-diode balanced mixer (BAM) with a local oscillator (LO). In this case, the local oscillator is an unmodulated copy of the laser lines incident on a second UTC-PD. A variable delay line is used to match the phase of the LO to the data stream. Insets depict optical, THz, and RF spectra at various points in the setup. AWG, arbitrary waveform generator; IM, intensity modulator; VOA, variable optical attenuator; UTC, uni-traveling-carrier photodiode; BAM, Schottky diode mixer; RTO, real-time oscilloscope. \textbf{b}, Resulting constellations for 16 and 32 QAM.}
\end{figure}

To benchmark the performance of our source, we wirelessly transmitted data while using an LO delivered via optical fiber. Figure \ref{80gbps} (a) shows a block diagram of the setup for the homodyne communication system used in the experiment. The output of the laser source was divided into two paths; 80\% of the output is used for the transmitter, and 20\% for the receiver. In the transmitter, the QAM data signal was upconverted to the intermediate frequency (IF, 12.5 GHz) with an arbitrary waveform generator (AWG) and applied to an optical intensity modulator (IM). The IQ signal was filtered by a Gaussian filter with a roll-off factor of 0.35. The modulated optical signal was incident on a UTC-PD to generate and radiate the 301 GHz signal from a 25-dBi horn antenna. In the receiver, a horn antenna with the same gain received the RF signal and was followed by a Schottky-barrier-diode fundamental balanced mixer (BAM). The mixer was pumped by the local oscillator (LO) signal with the same frequency, which was generated with a UTC-PD and amplified with an RF amplifier. The down-converted IF signal was amplified with a baseband amplifier and was demodulated and analyzed with a real time oscilloscope (RTO).

We evaluated the THz source using 16 and 32 QAM modulation formats. Fig. \ref{80gbps} (b) shows the maximum symbol rates attained within hard decision forward error correction (HD-FEC) limits. The HD-FEC limit results in rms EVM values sufficient for many applications as outlined in the IEEE 802.15.3d standard amendment. \cite{2017, Petrov2020} The maximum symbol rate achieved was 19 Gbaud with 16 QAM modulation, while it was 16 Gbaud with 32 QAM modulation. Total net data rates were 76 Gbps and 80 Gbps with 16 QAM and 32 QAM, respectively. Since higher SNR is required for higher-order modulation, the maximum symbol rate decreases from 16QAM to 32QAM. The difference in the symbol rate is mainly limited by the SNR required for each modulation format. 

To compare OIL with a more conventional amplification chain used in laser-based THz sources, we built the system shown in Fig.4 (a), which is based on electro-optic (EO) comb generation. A single mode, external cavity diode laser (ECDL) produced the primary comb line at 1550 nm. An optical phase modulator modulated the light at 37.5-GHz to produce the EO comb. By using an optical filter (OF), two wavelengths with a frequency difference of 300 GHz were selected. To obtain larger OSNR, a second optical filter with the same pass bands was used and optical amplifiers were added after each optical filter to compensate for loss. With this configuration, an optical output power of 100 mW with a OSNR ratio of 55 dB was achieved. 16 QAM data transmission was performed with the same setup shown in Fig. \ref{80gbps} \textbf{a}. Constellation diagrams for symbol rates of 5, 10, and 13 Gbaud are shown in Fig. 4. The maximum symbol rate within the HD-FEC limit was 13 Gbaud, which corresponds to the maximum date rate of 52 Gbps. The difference in the maximum symbol rates between two systems is likely due to the amount of the amplitude noise or the OSNR. This demonstrates that OIL, in general, is a convenient and cost-effective alternative to repeated filtering and amplification steps using waveshapers and EDFA's for low OSNR comb sources, as initially reported in. \cite{Albores-Mejia2015}

\begin{figure}[ht!]
\centering\includegraphics[width=3.3in]{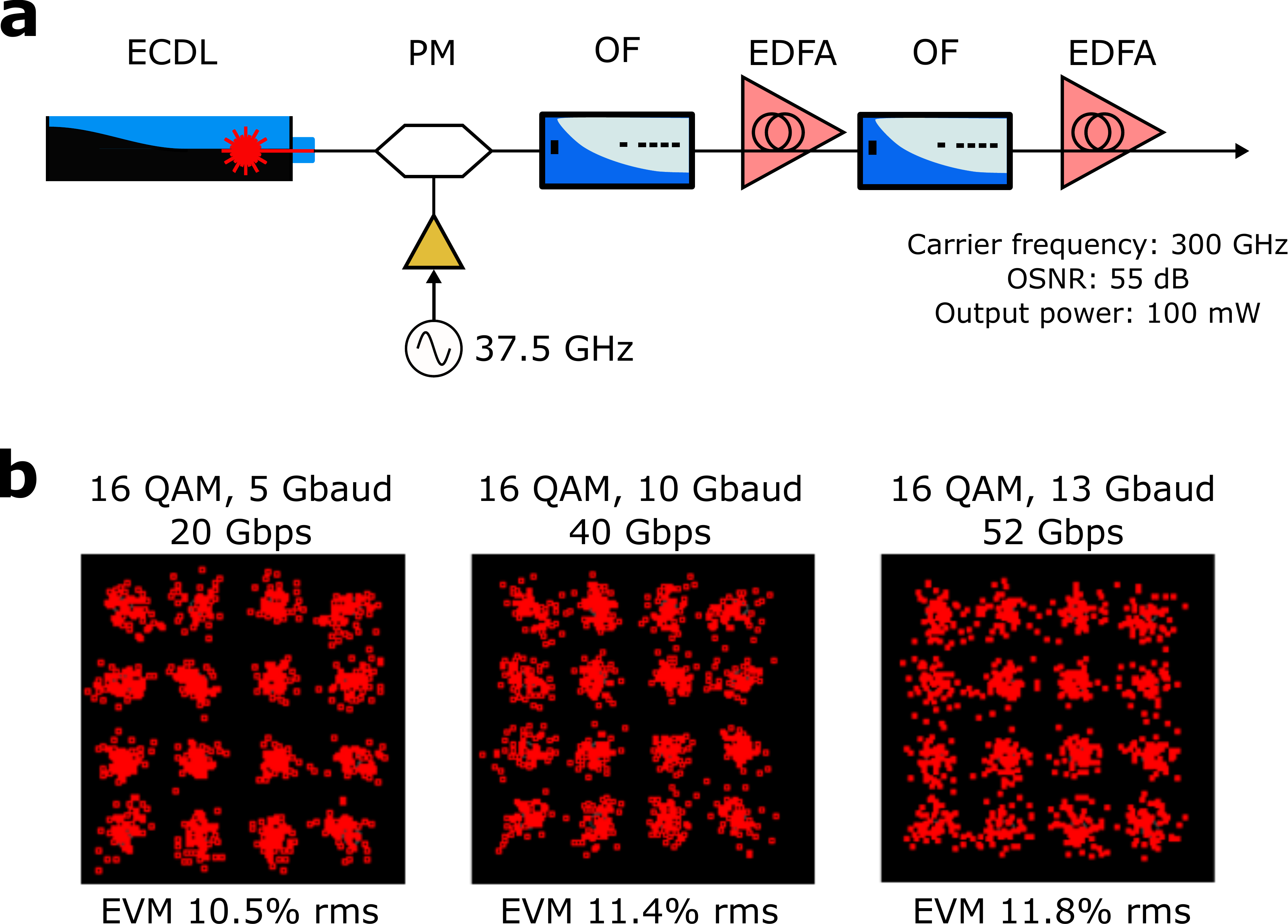}
\caption{\label{eocomb} \textbf{a}, EO-comb based laser system.  ECDL, external cavity diode laser; PM, phase modulator; EDFA, erbium-doped fiber amplifier; OF, optical filter. \textbf{b}, 16 QAM constellation diagrams with symbol rates of 5, 10, and 13 Gbaud.}
\end{figure}

\begin{figure}[ht!]
\centering\includegraphics[width=3.3in]{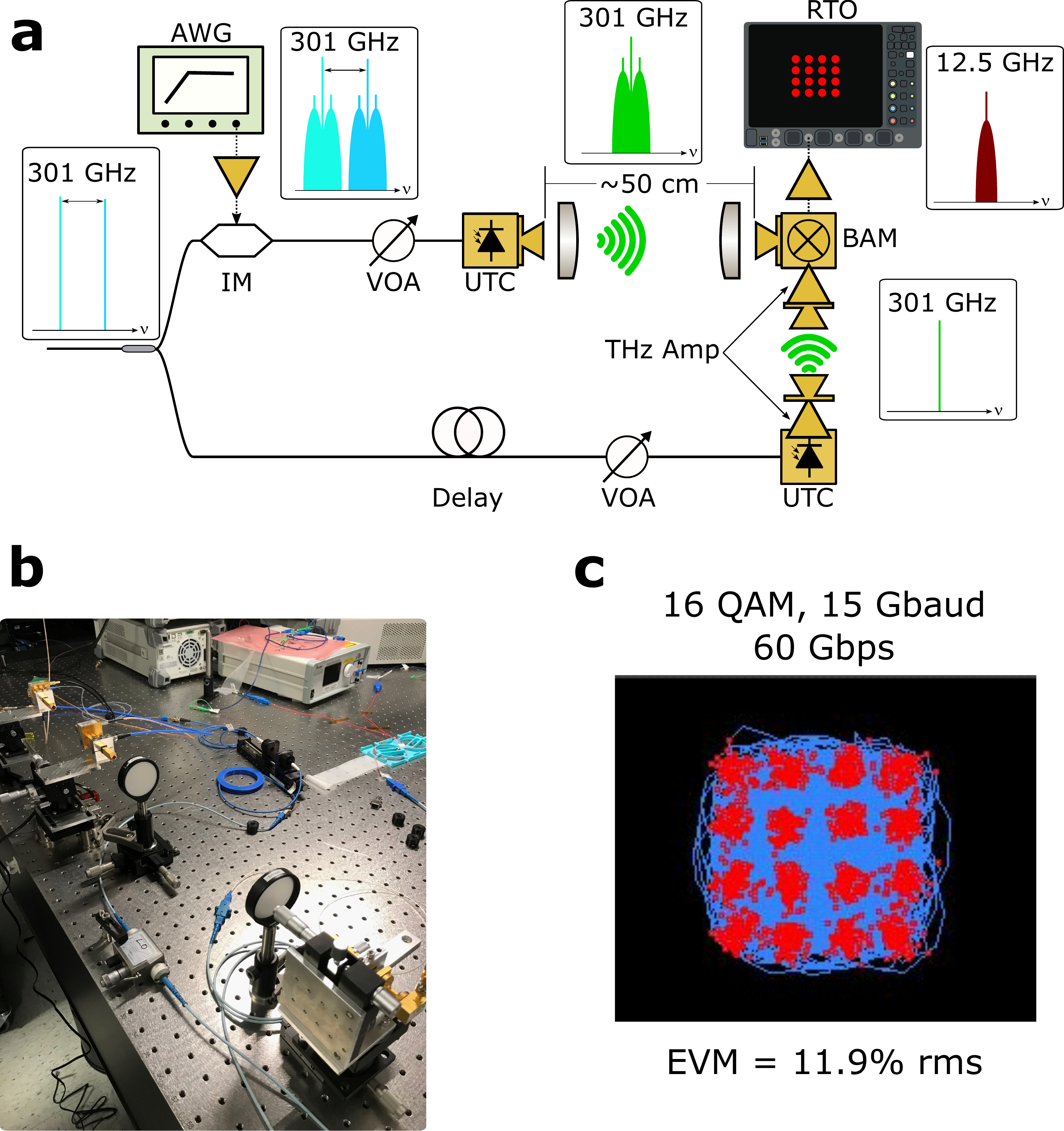}
\caption{\label{60gbps} \textbf{a}, Diagram of the fully wireless communication demonstration utilizing a Kerr microcomb. The source depicted in Fig. \ref{microwave} \textbf{a}, is split into two paths. In the first, the two laser lines are modulated using an intensity modulator driven by an AWG. The optically-carried information is attenuated before being down-converted on a UTC-PD and radiating from an antenna. A PTFE plano-convex lens collimates the radiation from the emitter, while a second PTFE lens focuses it on to a Schottky diode mixer. The distance between the transmitter and receiver is approximately 50 cm. An unaltered copy of the source is attenuated before impinging on a second UTC-PD. This transmits a THz clock signal which is mixed on the BAM. \textbf{b}, A picture of the setup. \textbf{c}, resulting constellation using 16 QAM with a 15 Gbaud rate. We transmit 60 Gbps with an rms EVM of 11.9\%}
\end{figure}

Finally, we conducted a transmission experiment using the Kerr-comb-based source by wirelessly delivering the LO and RF signals using the configuration shown in Fig. \ref{60gbps}. This configuration offers a realistic wireless link, which completely separates a transmitter and a receiver. The LO signal was amplified and emitted from a horn antenna. In the receiver, the LO signal was received by 25-dBi gain horn antenna, amplified again and fed to the LO port of the BAM. The distance between the RF ports of the transmitter (UTC-PD) and receiver (BAM) was set to 50 cm by collimating the THz wave with PTFE plano-convex lenses, while the distance between LO ports of the transmitter and receiver was 10 cm. In the experiment, we applied 16 QAM modulation. Fig. \ref{60gbps} (c) shows the constellation diagram with the maximum symbol rate of 15 Gbaud, or 60 Gbps. The EVM value was 11.9\%, which is within the HD-FEC limit. The degraded symbol rate compared with the case of a fiber connected LO (Fig. \ref{80gbps} (a)) is likely due to increased amplitude noise of the LO signal after the second amplification stage. 
 
\section{Discussion}

The data rates presented here are primarily limited by the maximum baud rate achievable with the AWG. The SNR at the receiver currently prevents us from using higher-order modulation formats. Future work will focus on increasing the data rate by using higher baud rate AWGs and exploring refinements to the THz transmitter for improved SNR. Moving forward, there are several routes available to improve the performance of the micro-comb based source itself. For instance, amplitude noise may be reduced by using back-facet injection of the comb light. In this way, injection light is not reflected by the output facet of the LD, which can interfere with the LD light upon photodetection. \cite{Liu2020a} While aspects of the source have demonstrable integrability, \cite{Shen2020, Jia2022, Balakier2014} a fully integrated hybrid photonic THz transmitter using a microcomb via photonic wire bonding or other methods has yet to be shown and will be an important milestone for future progress.

\section{Conclusions}
We have demonstrated that two laser diodes optically injection-locked to two teeth of a DKS microcomb produce a viable source for THz communications. This source leverages the inherent low-noise properties of a microcomb while improving the optical signal to noise by greater than 35 dB. Using homodyne detection, wireless transfer rates of 80 Gbps were achieved with a wired LO, while rates of 60 Gbps were demonstrated using simultaneous transfer of both the data and clock channels. The source can plausibly be integrated, offering a path forward for compact, precision THz sources. 

\begin{acknowledgments}
The authors would like to express sincere thanks to Keysight Technologies for providing equipment and technical support. Part of this research result was obtained from the commissioned research (No.00901) by National Institute of Information and Communications Technology (NICT), Japan.
\end{acknowledgments}

\section*{Author Declarations}
\subsection{Conflicts of Interest}
The authors have no conflicts to disclose.

\subsection{Author Contributions}
\textbf{Brendan M. Heffernan:} investigation (equal); visualization (lead); writing - original draft preparation (equal); writing - review and editing (equal). 
\textbf{Yuma Kawamoto:} investigation (equal), writing - review and editing (equal).
\textbf{Keisuke Maekawa:} investigation (equal), methodology (equal),  writing - original draft preparation (equal), writing - review and editing (equal).
\textbf{James Greenberg:} writing - review and editing (equal), visualization (supporting).
\textbf{Rubab Amin:} writing - review and editing (equal).
\textbf{Takashi Hori:} conceptualization (equal), project administration (equal).
\textbf{Tatsuya Tanigawa:} project administration (equal).
\textbf{Tadao Nagatsuma:} conceptualization (equal), funding acquisition (lead), methodology (equal), project administration (equal), supervision (equal), writing - review and editing (equal).
\textbf{Antoine Rolland:} conceptualization (equal), formal analysis (equal), methodology (equal), supervision (equal), writing - review and editing (equal).

\section*{Data Availability}
The data that support the findings of this study are available within the article.

\bibliography{refs}

\begin{thebibliography}{43}%
\makeatletter
\providecommand \@ifxundefined [1]{%
 \@ifx{#1\undefined}
}%
\providecommand \@ifnum [1]{%
 \ifnum #1\expandafter \@firstoftwo
 \else \expandafter \@secondoftwo
 \fi
}%
\providecommand \@ifx [1]{%
 \ifx #1\expandafter \@firstoftwo
 \else \expandafter \@secondoftwo
 \fi
}%
\providecommand \natexlab [1]{#1}%
\providecommand \enquote  [1]{``#1''}%
\providecommand \bibnamefont  [1]{#1}%
\providecommand \bibfnamefont [1]{#1}%
\providecommand \citenamefont [1]{#1}%
\providecommand \href@noop [0]{\@secondoftwo}%
\providecommand \href [0]{\begingroup \@sanitize@url \@href}%
\providecommand \@href[1]{\@@startlink{#1}\@@href}%
\providecommand \@@href[1]{\endgroup#1\@@endlink}%
\providecommand \@sanitize@url [0]{\catcode `\\12\catcode `\$12\catcode
  `\&12\catcode `\#12\catcode `\^12\catcode `\_12\catcode `\%12\relax}%
\providecommand \@@startlink[1]{}%
\providecommand \@@endlink[0]{}%
\providecommand \url  [0]{\begingroup\@sanitize@url \@url }%
\providecommand \@url [1]{\endgroup\@href {#1}{\urlprefix }}%
\providecommand \urlprefix  [0]{URL }%
\providecommand \Eprint [0]{\href }%
\providecommand \doibase [0]{http://dx.doi.org/}%
\providecommand \selectlanguage [0]{\@gobble}%
\providecommand \bibinfo  [0]{\@secondoftwo}%
\providecommand \bibfield  [0]{\@secondoftwo}%
\providecommand \translation [1]{[#1]}%
\providecommand \BibitemOpen [0]{}%
\providecommand \bibitemStop [0]{}%
\providecommand \bibitemNoStop [0]{.\EOS\space}%
\providecommand \EOS [0]{\spacefactor3000\relax}%
\providecommand \BibitemShut  [1]{\csname bibitem#1\endcsname}%
\let\auto@bib@innerbib\@empty
\bibitem [{\citenamefont {K{\"u}rner}, \citenamefont {Mittleman},\ and\
  \citenamefont {Nagatsuma}(2022)}]{Kurner2022}%
  \BibitemOpen
  \bibinfo {editor} {\bibfnamefont {T.}~\bibnamefont {K{\"u}rner}}, \bibinfo
  {editor} {\bibfnamefont {D.~M.}\ \bibnamefont {Mittleman}}, \ and\ \bibinfo
  {editor} {\bibfnamefont {T.}~\bibnamefont {Nagatsuma}},\ eds.,\ \href
  {\doibase 10.1007/978-3-030-73738-2} {\emph {\bibinfo {title} {{{THz
  Communications}}: Paving the Way towards Wireless Tbps}}},\ \bibinfo {series}
  {Springer Series in Optical Sciences}\ No.\ \bibinfo {number} {volume 234}\
  (\bibinfo  {publisher} {{Springer}},\ \bibinfo {address} {{Cham,
  Switzerland}},\ \bibinfo {year} {2022})\BibitemShut {NoStop}%
\bibitem [{\citenamefont {Dang}\ \emph {et~al.}(2020)\citenamefont {Dang},
  \citenamefont {Amin}, \citenamefont {Shihada},\ and\ \citenamefont
  {Alouini}}]{Dang2020}%
  \BibitemOpen
  \bibfield  {author} {\bibinfo {author} {\bibfnamefont {S.}~\bibnamefont
  {Dang}}, \bibinfo {author} {\bibfnamefont {O.}~\bibnamefont {Amin}}, \bibinfo
  {author} {\bibfnamefont {B.}~\bibnamefont {Shihada}}, \ and\ \bibinfo
  {author} {\bibfnamefont {M.-S.}\ \bibnamefont {Alouini}},\ }\bibfield
  {title} {\enquote {\bibinfo {title} {What should {{6G}} be?}}\ }\href
  {\doibase 10.1038/s41928-019-0355-6} {\bibfield  {journal} {\bibinfo
  {journal} {Nature Electronics}\ }\textbf {\bibinfo {volume} {3}},\ \bibinfo
  {pages} {20--29} (\bibinfo {year} {2020})}\BibitemShut {NoStop}%
\bibitem [{\citenamefont {Sengupta}, \citenamefont {Nagatsuma},\ and\
  \citenamefont {Mittleman}(2018)}]{Sengupta2018}%
  \BibitemOpen
  \bibfield  {author} {\bibinfo {author} {\bibfnamefont {K.}~\bibnamefont
  {Sengupta}}, \bibinfo {author} {\bibfnamefont {T.}~\bibnamefont {Nagatsuma}},
  \ and\ \bibinfo {author} {\bibfnamefont {D.~M.}\ \bibnamefont {Mittleman}},\
  }\bibfield  {title} {\enquote {\bibinfo {title} {Terahertz integrated
  electronic and hybrid electronic\textendash photonic systems},}\ }\href
  {\doibase 10.1038/s41928-018-0173-2} {\bibfield  {journal} {\bibinfo
  {journal} {Nature Electronics}\ }\textbf {\bibinfo {volume} {1}},\ \bibinfo
  {pages} {622--635} (\bibinfo {year} {2018})}\BibitemShut {NoStop}%
\bibitem [{\citenamefont {Pang}\ \emph {et~al.}(2022)\citenamefont {Pang},
  \citenamefont {Ozolins}, \citenamefont {Jia}, \citenamefont {Zhang},
  \citenamefont {Schatz}, \citenamefont {Udalcovs}, \citenamefont {Bobrovs},
  \citenamefont {Hu}, \citenamefont {Morioka}, \citenamefont {Sun},
  \citenamefont {Chen}, \citenamefont {Lourdudoss}, \citenamefont {Oxenloewe},
  \citenamefont {Popov},\ and\ \citenamefont {Yu}}]{Pang2022}%
  \BibitemOpen
  \bibfield  {author} {\bibinfo {author} {\bibfnamefont {X.}~\bibnamefont
  {Pang}}, \bibinfo {author} {\bibfnamefont {O.}~\bibnamefont {Ozolins}},
  \bibinfo {author} {\bibfnamefont {S.}~\bibnamefont {Jia}}, \bibinfo {author}
  {\bibfnamefont {L.}~\bibnamefont {Zhang}}, \bibinfo {author} {\bibfnamefont
  {R.}~\bibnamefont {Schatz}}, \bibinfo {author} {\bibfnamefont
  {A.}~\bibnamefont {Udalcovs}}, \bibinfo {author} {\bibfnamefont
  {V.}~\bibnamefont {Bobrovs}}, \bibinfo {author} {\bibfnamefont
  {H.}~\bibnamefont {Hu}}, \bibinfo {author} {\bibfnamefont {T.}~\bibnamefont
  {Morioka}}, \bibinfo {author} {\bibfnamefont {Y.}~\bibnamefont {Sun}},
  \bibinfo {author} {\bibfnamefont {J.}~\bibnamefont {Chen}}, \bibinfo {author}
  {\bibfnamefont {S.}~\bibnamefont {Lourdudoss}}, \bibinfo {author}
  {\bibfnamefont {L.~K.}\ \bibnamefont {Oxenloewe}}, \bibinfo {author}
  {\bibfnamefont {S.}~\bibnamefont {Popov}}, \ and\ \bibinfo {author}
  {\bibfnamefont {X.}~\bibnamefont {Yu}},\ }\bibfield  {title} {\enquote
  {\bibinfo {title} {Bridging the {{Terahertz Gap}}: {{Photonics-assisted
  Free-Space Communications}} from the {{Submillimeter-Wave}} to the
  {{Mid-Infrared}}},}\ }\href {\doibase 10.1109/JLT.2022.3153139} {\bibfield
  {journal} {\bibinfo  {journal} {Journal of Lightwave Technology}\ ,\ \bibinfo
  {pages} {1--1}} (\bibinfo {year} {2022})}\BibitemShut {NoStop}%
\bibitem [{\citenamefont {Yao}\ \emph {et~al.}(2022)\citenamefont {Yao},
  \citenamefont {Yao}, \citenamefont {Liu}, \citenamefont {Liu}, \citenamefont
  {Chen}, \citenamefont {Gong}, \citenamefont {Gong}, \citenamefont {Gong},
  \citenamefont {Yang}, \citenamefont {Yang}, \citenamefont {Yang},
  \citenamefont {Xiao}, \citenamefont {Xiao}, \citenamefont {Xiao},
  \citenamefont {Xiao},\ and\ \citenamefont {Xiao}}]{Yao2022}%
  \BibitemOpen
  \bibfield  {author} {\bibinfo {author} {\bibfnamefont {L.}~\bibnamefont
  {Yao}}, \bibinfo {author} {\bibfnamefont {L.}~\bibnamefont {Yao}}, \bibinfo
  {author} {\bibfnamefont {P.}~\bibnamefont {Liu}}, \bibinfo {author}
  {\bibfnamefont {P.}~\bibnamefont {Liu}}, \bibinfo {author} {\bibfnamefont
  {H.-J.}\ \bibnamefont {Chen}}, \bibinfo {author} {\bibfnamefont
  {Q.}~\bibnamefont {Gong}}, \bibinfo {author} {\bibfnamefont {Q.}~\bibnamefont
  {Gong}}, \bibinfo {author} {\bibfnamefont {Q.}~\bibnamefont {Gong}}, \bibinfo
  {author} {\bibfnamefont {Q.-F.}\ \bibnamefont {Yang}}, \bibinfo {author}
  {\bibfnamefont {Q.-F.}\ \bibnamefont {Yang}}, \bibinfo {author}
  {\bibfnamefont {Q.-F.}\ \bibnamefont {Yang}}, \bibinfo {author}
  {\bibfnamefont {Y.-F.}\ \bibnamefont {Xiao}}, \bibinfo {author}
  {\bibfnamefont {Y.-F.}\ \bibnamefont {Xiao}}, \bibinfo {author}
  {\bibfnamefont {Y.-F.}\ \bibnamefont {Xiao}}, \bibinfo {author}
  {\bibfnamefont {Y.-F.}\ \bibnamefont {Xiao}}, \ and\ \bibinfo {author}
  {\bibfnamefont {Y.-F.}\ \bibnamefont {Xiao}},\ }\bibfield  {title} {\enquote
  {\bibinfo {title} {Soliton microwave oscillators using oversized billion
  {{Q}} optical microresonators},}\ }\href {\doibase 10.1364/OPTICA.459130}
  {\bibfield  {journal} {\bibinfo  {journal} {Optica}\ }\textbf {\bibinfo
  {volume} {9}},\ \bibinfo {pages} {561--564} (\bibinfo {year}
  {2022})}\BibitemShut {NoStop}%
\bibitem [{\citenamefont {Jia}\ \emph {et~al.}(2022)\citenamefont {Jia},
  \citenamefont {Lo}, \citenamefont {Zhang}, \citenamefont {Ozolins},
  \citenamefont {Udalcovs}, \citenamefont {Kong}, \citenamefont {Pang},
  \citenamefont {Guzman}, \citenamefont {Yu}, \citenamefont {Xiao},
  \citenamefont {Popov}, \citenamefont {Chen}, \citenamefont {Carpintero},
  \citenamefont {Morioka}, \citenamefont {Hu},\ and\ \citenamefont
  {Oxenl{\o}we}}]{Jia2022}%
  \BibitemOpen
  \bibfield  {author} {\bibinfo {author} {\bibfnamefont {S.}~\bibnamefont
  {Jia}}, \bibinfo {author} {\bibfnamefont {M.-C.}\ \bibnamefont {Lo}},
  \bibinfo {author} {\bibfnamefont {L.}~\bibnamefont {Zhang}}, \bibinfo
  {author} {\bibfnamefont {O.}~\bibnamefont {Ozolins}}, \bibinfo {author}
  {\bibfnamefont {A.}~\bibnamefont {Udalcovs}}, \bibinfo {author}
  {\bibfnamefont {D.}~\bibnamefont {Kong}}, \bibinfo {author} {\bibfnamefont
  {X.}~\bibnamefont {Pang}}, \bibinfo {author} {\bibfnamefont {R.}~\bibnamefont
  {Guzman}}, \bibinfo {author} {\bibfnamefont {X.}~\bibnamefont {Yu}}, \bibinfo
  {author} {\bibfnamefont {S.}~\bibnamefont {Xiao}}, \bibinfo {author}
  {\bibfnamefont {S.}~\bibnamefont {Popov}}, \bibinfo {author} {\bibfnamefont
  {J.}~\bibnamefont {Chen}}, \bibinfo {author} {\bibfnamefont {G.}~\bibnamefont
  {Carpintero}}, \bibinfo {author} {\bibfnamefont {T.}~\bibnamefont {Morioka}},
  \bibinfo {author} {\bibfnamefont {H.}~\bibnamefont {Hu}}, \ and\ \bibinfo
  {author} {\bibfnamefont {L.~K.}\ \bibnamefont {Oxenl{\o}we}},\ }\bibfield
  {title} {\enquote {\bibinfo {title} {Integrated dual-laser photonic chip for
  high-purity carrier generation enabling ultrafast terahertz wireless
  communications},}\ }\href {\doibase 10.1038/s41467-022-29049-2} {\bibfield
  {journal} {\bibinfo  {journal} {Nature Communications}\ }\textbf {\bibinfo
  {volume} {13}},\ \bibinfo {pages} {1388} (\bibinfo {year}
  {2022})}\BibitemShut {NoStop}%
\bibitem [{\citenamefont {Safian}, \citenamefont {Ghazi},\ and\ \citenamefont
  {Mohammadian}(2019)}]{Safian2019}%
  \BibitemOpen
  \bibfield  {author} {\bibinfo {author} {\bibfnamefont {R.}~\bibnamefont
  {Safian}}, \bibinfo {author} {\bibfnamefont {G.}~\bibnamefont {Ghazi}}, \
  and\ \bibinfo {author} {\bibfnamefont {N.}~\bibnamefont {Mohammadian}},\
  }\bibfield  {title} {\enquote {\bibinfo {title} {Review of photomixing
  continuous-wave terahertz systems and current application trends in terahertz
  domain},}\ }\href {\doibase 10.1117/1.OE.58.11.110901} {\bibfield  {journal}
  {\bibinfo  {journal} {Optical Engineering}\ }\textbf {\bibinfo {volume}
  {58}},\ \bibinfo {pages} {110901} (\bibinfo {year} {2019})}\BibitemShut
  {NoStop}%
\bibitem [{\citenamefont {Ummethala}\ \emph {et~al.}(2019)\citenamefont
  {Ummethala}, \citenamefont {Harter}, \citenamefont {Koehnle}, \citenamefont
  {Li}, \citenamefont {Muehlbrandt}, \citenamefont {Kutuvantavida},
  \citenamefont {Kemal}, \citenamefont {{Marin-Palomo}}, \citenamefont
  {Schaefer}, \citenamefont {Tessmann}, \citenamefont {Garlapati},
  \citenamefont {Bacher}, \citenamefont {Hahn}, \citenamefont {Walther},
  \citenamefont {Zwick}, \citenamefont {Randel}, \citenamefont {Freude},\ and\
  \citenamefont {Koos}}]{Ummethala2019}%
  \BibitemOpen
  \bibfield  {author} {\bibinfo {author} {\bibfnamefont {S.}~\bibnamefont
  {Ummethala}}, \bibinfo {author} {\bibfnamefont {T.}~\bibnamefont {Harter}},
  \bibinfo {author} {\bibfnamefont {K.}~\bibnamefont {Koehnle}}, \bibinfo
  {author} {\bibfnamefont {Z.}~\bibnamefont {Li}}, \bibinfo {author}
  {\bibfnamefont {S.}~\bibnamefont {Muehlbrandt}}, \bibinfo {author}
  {\bibfnamefont {Y.}~\bibnamefont {Kutuvantavida}}, \bibinfo {author}
  {\bibfnamefont {J.}~\bibnamefont {Kemal}}, \bibinfo {author} {\bibfnamefont
  {P.}~\bibnamefont {{Marin-Palomo}}}, \bibinfo {author} {\bibfnamefont
  {J.}~\bibnamefont {Schaefer}}, \bibinfo {author} {\bibfnamefont
  {A.}~\bibnamefont {Tessmann}}, \bibinfo {author} {\bibfnamefont {S.~K.}\
  \bibnamefont {Garlapati}}, \bibinfo {author} {\bibfnamefont {A.}~\bibnamefont
  {Bacher}}, \bibinfo {author} {\bibfnamefont {L.}~\bibnamefont {Hahn}},
  \bibinfo {author} {\bibfnamefont {M.}~\bibnamefont {Walther}}, \bibinfo
  {author} {\bibfnamefont {T.}~\bibnamefont {Zwick}}, \bibinfo {author}
  {\bibfnamefont {S.}~\bibnamefont {Randel}}, \bibinfo {author} {\bibfnamefont
  {W.}~\bibnamefont {Freude}}, \ and\ \bibinfo {author} {\bibfnamefont
  {C.}~\bibnamefont {Koos}},\ }\bibfield  {title} {\enquote {\bibinfo {title}
  {{{THz-to-optical}} conversion in wireless communications using an
  ultra-broadband plasmonic modulator},}\ }\href {\doibase
  10.1038/s41566-019-0475-6} {\bibfield  {journal} {\bibinfo  {journal} {Nature
  Photonics}\ }\textbf {\bibinfo {volume} {13}},\ \bibinfo {pages} {519--524}
  (\bibinfo {year} {2019})}\BibitemShut {NoStop}%
\bibitem [{\citenamefont {Fortier}\ \emph {et~al.}(2016)\citenamefont
  {Fortier}, \citenamefont {Rolland}, \citenamefont {Quinlan}, \citenamefont
  {Baynes}, \citenamefont {Metcalf}, \citenamefont {Hati}, \citenamefont
  {Ludlow}, \citenamefont {Hinkley}, \citenamefont {Shimizu}, \citenamefont
  {Ishibashi}, \citenamefont {Campbell},\ and\ \citenamefont
  {Diddams}}]{Fortier2016}%
  \BibitemOpen
  \bibfield  {author} {\bibinfo {author} {\bibfnamefont {T.~M.}\ \bibnamefont
  {Fortier}}, \bibinfo {author} {\bibfnamefont {A.}~\bibnamefont {Rolland}},
  \bibinfo {author} {\bibfnamefont {F.}~\bibnamefont {Quinlan}}, \bibinfo
  {author} {\bibfnamefont {F.~N.}\ \bibnamefont {Baynes}}, \bibinfo {author}
  {\bibfnamefont {A.~J.}\ \bibnamefont {Metcalf}}, \bibinfo {author}
  {\bibfnamefont {A.}~\bibnamefont {Hati}}, \bibinfo {author} {\bibfnamefont
  {A.~D.}\ \bibnamefont {Ludlow}}, \bibinfo {author} {\bibfnamefont
  {N.}~\bibnamefont {Hinkley}}, \bibinfo {author} {\bibfnamefont
  {M.}~\bibnamefont {Shimizu}}, \bibinfo {author} {\bibfnamefont
  {T.}~\bibnamefont {Ishibashi}}, \bibinfo {author} {\bibfnamefont {J.~C.}\
  \bibnamefont {Campbell}}, \ and\ \bibinfo {author} {\bibfnamefont {S.~A.}\
  \bibnamefont {Diddams}},\ }\bibfield  {title} {\enquote {\bibinfo {title}
  {Optically referenced broadband electronic synthesizer with 15 digits of
  resolution},}\ }\href {\doibase 10.1002/lpor.201500307} {\bibfield  {journal}
  {\bibinfo  {journal} {Laser \& Photonics Reviews}\ }\textbf {\bibinfo
  {volume} {10}},\ \bibinfo {pages} {780--790} (\bibinfo {year}
  {2016})}\BibitemShut {NoStop}%
\bibitem [{\citenamefont {Xie}\ \emph {et~al.}(2017)\citenamefont {Xie},
  \citenamefont {Bouchand}, \citenamefont {Nicolodi}, \citenamefont {Giunta},
  \citenamefont {H{\"a}nsel}, \citenamefont {Lezius}, \citenamefont {Joshi},
  \citenamefont {Datta}, \citenamefont {Alexandre}, \citenamefont {Lours},
  \citenamefont {Tremblin}, \citenamefont {Santarelli}, \citenamefont
  {Holzwarth},\ and\ \citenamefont {Le~Coq}}]{Xie2017}%
  \BibitemOpen
  \bibfield  {author} {\bibinfo {author} {\bibfnamefont {X.}~\bibnamefont
  {Xie}}, \bibinfo {author} {\bibfnamefont {R.}~\bibnamefont {Bouchand}},
  \bibinfo {author} {\bibfnamefont {D.}~\bibnamefont {Nicolodi}}, \bibinfo
  {author} {\bibfnamefont {M.}~\bibnamefont {Giunta}}, \bibinfo {author}
  {\bibfnamefont {W.}~\bibnamefont {H{\"a}nsel}}, \bibinfo {author}
  {\bibfnamefont {M.}~\bibnamefont {Lezius}}, \bibinfo {author} {\bibfnamefont
  {A.}~\bibnamefont {Joshi}}, \bibinfo {author} {\bibfnamefont
  {S.}~\bibnamefont {Datta}}, \bibinfo {author} {\bibfnamefont
  {C.}~\bibnamefont {Alexandre}}, \bibinfo {author} {\bibfnamefont
  {M.}~\bibnamefont {Lours}}, \bibinfo {author} {\bibfnamefont {P.-A.}\
  \bibnamefont {Tremblin}}, \bibinfo {author} {\bibfnamefont {G.}~\bibnamefont
  {Santarelli}}, \bibinfo {author} {\bibfnamefont {R.}~\bibnamefont
  {Holzwarth}}, \ and\ \bibinfo {author} {\bibfnamefont {Y.}~\bibnamefont
  {Le~Coq}},\ }\bibfield  {title} {\enquote {\bibinfo {title} {Photonic
  microwave signals with zeptosecond-level absolute timing noise},}\ }\href
  {\doibase 10.1038/nphoton.2016.215} {\bibfield  {journal} {\bibinfo
  {journal} {Nature Photonics}\ }\textbf {\bibinfo {volume} {11}},\ \bibinfo
  {pages} {44--47} (\bibinfo {year} {2017})}\BibitemShut {NoStop}%
\bibitem [{\citenamefont {Nakamura}\ \emph {et~al.}(2020)\citenamefont
  {Nakamura}, \citenamefont {{Davila-Rodriguez}}, \citenamefont {Leopardi},
  \citenamefont {Sherman}, \citenamefont {Fortier}, \citenamefont {Xie},
  \citenamefont {Campbell}, \citenamefont {McGrew}, \citenamefont {Zhang},
  \citenamefont {Hassan}, \citenamefont {Nicolodi}, \citenamefont {Beloy},
  \citenamefont {Ludlow}, \citenamefont {Diddams},\ and\ \citenamefont
  {Quinlan}}]{Nakamura2020}%
  \BibitemOpen
  \bibfield  {author} {\bibinfo {author} {\bibfnamefont {T.}~\bibnamefont
  {Nakamura}}, \bibinfo {author} {\bibfnamefont {J.}~\bibnamefont
  {{Davila-Rodriguez}}}, \bibinfo {author} {\bibfnamefont {H.}~\bibnamefont
  {Leopardi}}, \bibinfo {author} {\bibfnamefont {J.~A.}\ \bibnamefont
  {Sherman}}, \bibinfo {author} {\bibfnamefont {T.~M.}\ \bibnamefont
  {Fortier}}, \bibinfo {author} {\bibfnamefont {X.}~\bibnamefont {Xie}},
  \bibinfo {author} {\bibfnamefont {J.~C.}\ \bibnamefont {Campbell}}, \bibinfo
  {author} {\bibfnamefont {W.~F.}\ \bibnamefont {McGrew}}, \bibinfo {author}
  {\bibfnamefont {X.}~\bibnamefont {Zhang}}, \bibinfo {author} {\bibfnamefont
  {Y.~S.}\ \bibnamefont {Hassan}}, \bibinfo {author} {\bibfnamefont
  {D.}~\bibnamefont {Nicolodi}}, \bibinfo {author} {\bibfnamefont
  {K.}~\bibnamefont {Beloy}}, \bibinfo {author} {\bibfnamefont {A.~D.}\
  \bibnamefont {Ludlow}}, \bibinfo {author} {\bibfnamefont {S.~A.}\
  \bibnamefont {Diddams}}, \ and\ \bibinfo {author} {\bibfnamefont
  {F.}~\bibnamefont {Quinlan}},\ }\bibfield  {title} {\enquote {\bibinfo
  {title} {Coherent optical clock down-conversion for microwave frequencies
  with 10-18 instability},}\ }\href {\doibase 10.1126/science.abb2473}
  {\bibfield  {journal} {\bibinfo  {journal} {Science}\ }\textbf {\bibinfo
  {volume} {368}},\ \bibinfo {pages} {889--892} (\bibinfo {year}
  {2020})}\BibitemShut {NoStop}%
\bibitem [{\citenamefont {Tetsumoto}\ \emph {et~al.}(2021)\citenamefont
  {Tetsumoto}, \citenamefont {Nagatsuma}, \citenamefont {Fermann},
  \citenamefont {Navickaite}, \citenamefont {Geiselmann},\ and\ \citenamefont
  {Rolland}}]{Tetsumoto2021}%
  \BibitemOpen
  \bibfield  {author} {\bibinfo {author} {\bibfnamefont {T.}~\bibnamefont
  {Tetsumoto}}, \bibinfo {author} {\bibfnamefont {T.}~\bibnamefont
  {Nagatsuma}}, \bibinfo {author} {\bibfnamefont {M.~E.}\ \bibnamefont
  {Fermann}}, \bibinfo {author} {\bibfnamefont {G.}~\bibnamefont {Navickaite}},
  \bibinfo {author} {\bibfnamefont {M.}~\bibnamefont {Geiselmann}}, \ and\
  \bibinfo {author} {\bibfnamefont {A.}~\bibnamefont {Rolland}},\ }\bibfield
  {title} {\enquote {\bibinfo {title} {Optically referenced 300 {{GHz}}
  millimetre-wave oscillator},}\ }\href {\doibase 10.1038/s41566-021-00790-2}
  {\bibfield  {journal} {\bibinfo  {journal} {Nature Photonics}\ }\textbf
  {\bibinfo {volume} {15}},\ \bibinfo {pages} {516--522} (\bibinfo {year}
  {2021})}\BibitemShut {NoStop}%
\bibitem [{\citenamefont {Chen}\ \emph {et~al.}(2018)\citenamefont {Chen},
  \citenamefont {Kuylenstierna}, \citenamefont {Gunnarsson}, \citenamefont
  {He}, \citenamefont {Eriksson}, \citenamefont {Swahn},\ and\ \citenamefont
  {Zirath}}]{Chen2018}%
  \BibitemOpen
  \bibfield  {author} {\bibinfo {author} {\bibfnamefont {J.}~\bibnamefont
  {Chen}}, \bibinfo {author} {\bibfnamefont {D.}~\bibnamefont {Kuylenstierna}},
  \bibinfo {author} {\bibfnamefont {S.~E.}\ \bibnamefont {Gunnarsson}},
  \bibinfo {author} {\bibfnamefont {Z.~S.}\ \bibnamefont {He}}, \bibinfo
  {author} {\bibfnamefont {T.}~\bibnamefont {Eriksson}}, \bibinfo {author}
  {\bibfnamefont {T.}~\bibnamefont {Swahn}}, \ and\ \bibinfo {author}
  {\bibfnamefont {H.}~\bibnamefont {Zirath}},\ }\bibfield  {title} {\enquote
  {\bibinfo {title} {Influence of {{White LO Noise}} on {{Wideband
  Communication}}},}\ }\href {\doibase 10.1109/TMTT.2018.2814040} {\bibfield
  {journal} {\bibinfo  {journal} {IEEE Transactions on Microwave Theory and
  Techniques}\ }\textbf {\bibinfo {volume} {66}},\ \bibinfo {pages}
  {3349--3359} (\bibinfo {year} {2018})}\BibitemShut {NoStop}%
\bibitem [{\citenamefont {Pang}\ \emph {et~al.}(2011)\citenamefont {Pang},
  \citenamefont {Caballero}, \citenamefont {Dogadaev}, \citenamefont {Arlunno},
  \citenamefont {Borkowski}, \citenamefont {Pedersen}, \citenamefont {Deng},
  \citenamefont {Karinou}, \citenamefont {Roubeau}, \citenamefont {Zibar},
  \citenamefont {Yu},\ and\ \citenamefont {Monroy}}]{Pang2011}%
  \BibitemOpen
  \bibfield  {author} {\bibinfo {author} {\bibfnamefont {X.}~\bibnamefont
  {Pang}}, \bibinfo {author} {\bibfnamefont {A.}~\bibnamefont {Caballero}},
  \bibinfo {author} {\bibfnamefont {A.}~\bibnamefont {Dogadaev}}, \bibinfo
  {author} {\bibfnamefont {V.}~\bibnamefont {Arlunno}}, \bibinfo {author}
  {\bibfnamefont {R.}~\bibnamefont {Borkowski}}, \bibinfo {author}
  {\bibfnamefont {J.~S.}\ \bibnamefont {Pedersen}}, \bibinfo {author}
  {\bibfnamefont {L.}~\bibnamefont {Deng}}, \bibinfo {author} {\bibfnamefont
  {F.}~\bibnamefont {Karinou}}, \bibinfo {author} {\bibfnamefont
  {F.}~\bibnamefont {Roubeau}}, \bibinfo {author} {\bibfnamefont
  {D.}~\bibnamefont {Zibar}}, \bibinfo {author} {\bibfnamefont
  {X.}~\bibnamefont {Yu}}, \ and\ \bibinfo {author} {\bibfnamefont {I.~T.}\
  \bibnamefont {Monroy}},\ }\bibfield  {title} {\enquote {\bibinfo {title} {100
  {{Gbit}}/s hybrid optical fiber-wireless link in the {{W-band}}
  (75\textendash 110 {{GHz}})},}\ }\href {\doibase 10.1364/OE.19.024944}
  {\bibfield  {journal} {\bibinfo  {journal} {Optics Express}\ }\textbf
  {\bibinfo {volume} {19}},\ \bibinfo {pages} {24944--24949} (\bibinfo {year}
  {2011})}\BibitemShut {NoStop}%
\bibitem [{\citenamefont {Nagatsuma}\ \emph {et~al.}(2013)\citenamefont
  {Nagatsuma}, \citenamefont {Horiguchi}, \citenamefont {Minamikata},
  \citenamefont {Yoshimizu}, \citenamefont {Hisatake}, \citenamefont {Kuwano},
  \citenamefont {Yoshimoto}, \citenamefont {Terada},\ and\ \citenamefont
  {Takahashi}}]{Nagatsuma2013}%
  \BibitemOpen
  \bibfield  {author} {\bibinfo {author} {\bibfnamefont {T.}~\bibnamefont
  {Nagatsuma}}, \bibinfo {author} {\bibfnamefont {S.}~\bibnamefont
  {Horiguchi}}, \bibinfo {author} {\bibfnamefont {Y.}~\bibnamefont
  {Minamikata}}, \bibinfo {author} {\bibfnamefont {Y.}~\bibnamefont
  {Yoshimizu}}, \bibinfo {author} {\bibfnamefont {S.}~\bibnamefont {Hisatake}},
  \bibinfo {author} {\bibfnamefont {S.}~\bibnamefont {Kuwano}}, \bibinfo
  {author} {\bibfnamefont {N.}~\bibnamefont {Yoshimoto}}, \bibinfo {author}
  {\bibfnamefont {J.}~\bibnamefont {Terada}}, \ and\ \bibinfo {author}
  {\bibfnamefont {H.}~\bibnamefont {Takahashi}},\ }\bibfield  {title} {\enquote
  {\bibinfo {title} {Terahertz wireless communications based on photonics
  technologies},}\ }\href {\doibase 10.1364/OE.21.023736} {\bibfield  {journal}
  {\bibinfo  {journal} {Optics Express}\ }\textbf {\bibinfo {volume} {21}},\
  \bibinfo {pages} {23736--23747} (\bibinfo {year} {2013})}\BibitemShut
  {NoStop}%
\bibitem [{\citenamefont {Harter}\ \emph {et~al.}(2020)\citenamefont {Harter},
  \citenamefont {F{\"u}llner}, \citenamefont {Kemal}, \citenamefont
  {Ummethala}, \citenamefont {Steinmann}, \citenamefont {Brosi}, \citenamefont
  {Hesler}, \citenamefont {Br{\"u}ndermann}, \citenamefont {M{\"u}ller},
  \citenamefont {Freude}, \citenamefont {Randel},\ and\ \citenamefont
  {Koos}}]{Harter2020}%
  \BibitemOpen
  \bibfield  {author} {\bibinfo {author} {\bibfnamefont {T.}~\bibnamefont
  {Harter}}, \bibinfo {author} {\bibfnamefont {C.}~\bibnamefont {F{\"u}llner}},
  \bibinfo {author} {\bibfnamefont {J.~N.}\ \bibnamefont {Kemal}}, \bibinfo
  {author} {\bibfnamefont {S.}~\bibnamefont {Ummethala}}, \bibinfo {author}
  {\bibfnamefont {J.~L.}\ \bibnamefont {Steinmann}}, \bibinfo {author}
  {\bibfnamefont {M.}~\bibnamefont {Brosi}}, \bibinfo {author} {\bibfnamefont
  {J.~L.}\ \bibnamefont {Hesler}}, \bibinfo {author} {\bibfnamefont
  {E.}~\bibnamefont {Br{\"u}ndermann}}, \bibinfo {author} {\bibfnamefont
  {A.-S.}\ \bibnamefont {M{\"u}ller}}, \bibinfo {author} {\bibfnamefont
  {W.}~\bibnamefont {Freude}}, \bibinfo {author} {\bibfnamefont
  {S.}~\bibnamefont {Randel}}, \ and\ \bibinfo {author} {\bibfnamefont
  {C.}~\bibnamefont {Koos}},\ }\bibfield  {title} {\enquote {\bibinfo {title}
  {Generalized {{Kramers}}\textendash{{Kronig}} receiver for coherent terahertz
  communications},}\ }\href {\doibase 10.1038/s41566-020-0675-0} {\bibfield
  {journal} {\bibinfo  {journal} {Nature Photonics}\ }\textbf {\bibinfo
  {volume} {14}},\ \bibinfo {pages} {601--606} (\bibinfo {year}
  {2020})}\BibitemShut {NoStop}%
\bibitem [{\citenamefont {Kanno}\ \emph {et~al.}(2011)\citenamefont {Kanno},
  \citenamefont {Inagaki}, \citenamefont {Morohashi}, \citenamefont {Sakamoto},
  \citenamefont {Kuri}, \citenamefont {Hosako}, \citenamefont {Kawanishi},
  \citenamefont {Yoshida},\ and\ \citenamefont {Kitayama}}]{Kanno2011}%
  \BibitemOpen
  \bibfield  {author} {\bibinfo {author} {\bibfnamefont {A.}~\bibnamefont
  {Kanno}}, \bibinfo {author} {\bibfnamefont {K.}~\bibnamefont {Inagaki}},
  \bibinfo {author} {\bibfnamefont {I.}~\bibnamefont {Morohashi}}, \bibinfo
  {author} {\bibfnamefont {T.}~\bibnamefont {Sakamoto}}, \bibinfo {author}
  {\bibfnamefont {T.}~\bibnamefont {Kuri}}, \bibinfo {author} {\bibfnamefont
  {I.}~\bibnamefont {Hosako}}, \bibinfo {author} {\bibfnamefont
  {T.}~\bibnamefont {Kawanishi}}, \bibinfo {author} {\bibfnamefont
  {Y.}~\bibnamefont {Yoshida}}, \ and\ \bibinfo {author} {\bibfnamefont
  {K.-i.}\ \bibnamefont {Kitayama}},\ }\bibfield  {title} {\enquote {\bibinfo
  {title} {40 {{Gb}}/s {{W-band}} (75\textendash 110 {{GHz}}) 16-{{QAM}}
  radio-over-fiber signal generation and its wireless transmission},}\ }\href
  {\doibase 10.1364/OE.19.000B56} {\bibfield  {journal} {\bibinfo  {journal}
  {Optics Express}\ }\textbf {\bibinfo {volume} {19}},\ \bibinfo {pages}
  {B56--B63} (\bibinfo {year} {2011})}\BibitemShut {NoStop}%
\bibitem [{\citenamefont {Koenig}\ \emph {et~al.}(2013)\citenamefont {Koenig},
  \citenamefont {{Lopez-Diaz}}, \citenamefont {Antes}, \citenamefont {Boes},
  \citenamefont {Henneberger}, \citenamefont {Leuther}, \citenamefont
  {Tessmann}, \citenamefont {Schmogrow}, \citenamefont {Hillerkuss},
  \citenamefont {Palmer}, \citenamefont {Zwick}, \citenamefont {Koos},
  \citenamefont {Freude}, \citenamefont {Ambacher}, \citenamefont {Leuthold},\
  and\ \citenamefont {Kallfass}}]{Koenig2013}%
  \BibitemOpen
  \bibfield  {author} {\bibinfo {author} {\bibfnamefont {S.}~\bibnamefont
  {Koenig}}, \bibinfo {author} {\bibfnamefont {D.}~\bibnamefont
  {{Lopez-Diaz}}}, \bibinfo {author} {\bibfnamefont {J.}~\bibnamefont {Antes}},
  \bibinfo {author} {\bibfnamefont {F.}~\bibnamefont {Boes}}, \bibinfo {author}
  {\bibfnamefont {R.}~\bibnamefont {Henneberger}}, \bibinfo {author}
  {\bibfnamefont {A.}~\bibnamefont {Leuther}}, \bibinfo {author} {\bibfnamefont
  {A.}~\bibnamefont {Tessmann}}, \bibinfo {author} {\bibfnamefont
  {R.}~\bibnamefont {Schmogrow}}, \bibinfo {author} {\bibfnamefont
  {D.}~\bibnamefont {Hillerkuss}}, \bibinfo {author} {\bibfnamefont
  {R.}~\bibnamefont {Palmer}}, \bibinfo {author} {\bibfnamefont
  {T.}~\bibnamefont {Zwick}}, \bibinfo {author} {\bibfnamefont
  {C.}~\bibnamefont {Koos}}, \bibinfo {author} {\bibfnamefont {W.}~\bibnamefont
  {Freude}}, \bibinfo {author} {\bibfnamefont {O.}~\bibnamefont {Ambacher}},
  \bibinfo {author} {\bibfnamefont {J.}~\bibnamefont {Leuthold}}, \ and\
  \bibinfo {author} {\bibfnamefont {I.}~\bibnamefont {Kallfass}},\ }\bibfield
  {title} {\enquote {\bibinfo {title} {Wireless sub-{{THz}} communication
  system with high data rate},}\ }\href {\doibase 10.1038/nphoton.2013.275}
  {\bibfield  {journal} {\bibinfo  {journal} {Nature Photonics}\ }\textbf
  {\bibinfo {volume} {7}},\ \bibinfo {pages} {977--981} (\bibinfo {year}
  {2013})}\BibitemShut {NoStop}%
\bibitem [{\citenamefont {Mohammad}\ \emph {et~al.}(2018)\citenamefont
  {Mohammad}, \citenamefont {Shams}, \citenamefont {Liu}, \citenamefont
  {Graham}, \citenamefont {Natrella}, \citenamefont {Seeds},\ and\
  \citenamefont {Renaud}}]{Mohammad2018a}%
  \BibitemOpen
  \bibfield  {author} {\bibinfo {author} {\bibfnamefont {A.~W.}\ \bibnamefont
  {Mohammad}}, \bibinfo {author} {\bibfnamefont {H.}~\bibnamefont {Shams}},
  \bibinfo {author} {\bibfnamefont {C.-P.}\ \bibnamefont {Liu}}, \bibinfo
  {author} {\bibfnamefont {C.}~\bibnamefont {Graham}}, \bibinfo {author}
  {\bibfnamefont {M.}~\bibnamefont {Natrella}}, \bibinfo {author}
  {\bibfnamefont {A.~J.}\ \bibnamefont {Seeds}}, \ and\ \bibinfo {author}
  {\bibfnamefont {C.~C.}\ \bibnamefont {Renaud}},\ }\bibfield  {title}
  {\enquote {\bibinfo {title} {60-{{GHz Transmission Link Using Uni-Traveling
  Carrier Photodiodes}} at the {{Transmitter}} and the {{Receiver}}},}\ }\href
  {\doibase 10.1109/JLT.2018.2849938} {\bibfield  {journal} {\bibinfo
  {journal} {Journal of Lightwave Technology}\ }\textbf {\bibinfo {volume}
  {36}},\ \bibinfo {pages} {4507--4513} (\bibinfo {year} {2018})}\BibitemShut
  {NoStop}%
\bibitem [{\citenamefont {Liu}\ \emph {et~al.}(2022)\citenamefont {Liu},
  \citenamefont {Qiu}, \citenamefont {Ji}, \citenamefont {He}, \citenamefont
  {Riemensberger}, \citenamefont {Hafermann}, \citenamefont {Wang},
  \citenamefont {Liu}, \citenamefont {Ronning},\ and\ \citenamefont
  {Kippenberg}}]{Liu2022}%
  \BibitemOpen
  \bibfield  {author} {\bibinfo {author} {\bibfnamefont {Y.}~\bibnamefont
  {Liu}}, \bibinfo {author} {\bibfnamefont {Z.}~\bibnamefont {Qiu}}, \bibinfo
  {author} {\bibfnamefont {X.}~\bibnamefont {Ji}}, \bibinfo {author}
  {\bibfnamefont {J.}~\bibnamefont {He}}, \bibinfo {author} {\bibfnamefont
  {J.}~\bibnamefont {Riemensberger}}, \bibinfo {author} {\bibfnamefont
  {M.}~\bibnamefont {Hafermann}}, \bibinfo {author} {\bibfnamefont {R.~N.}\
  \bibnamefont {Wang}}, \bibinfo {author} {\bibfnamefont {J.}~\bibnamefont
  {Liu}}, \bibinfo {author} {\bibfnamefont {C.}~\bibnamefont {Ronning}}, \ and\
  \bibinfo {author} {\bibfnamefont {T.~J.}\ \bibnamefont {Kippenberg}},\
  }\bibfield  {title} {\enquote {\bibinfo {title} {A photonic integrated
  circuit based erbium-doped amplifier},}\ }\href@noop {} {\bibfield  {journal}
  {\bibinfo  {journal} {arXiv:2204.02202 [physics]}\ } (\bibinfo {year}
  {2022})},\ \Eprint {http://arxiv.org/abs/2204.02202} {arXiv:2204.02202
  [physics]} \BibitemShut {NoStop}%
\bibitem [{\citenamefont {Liu}\ and\ \citenamefont
  {Slav{\'i}k}(2020)}]{Liu2020a}%
  \BibitemOpen
  \bibfield  {author} {\bibinfo {author} {\bibfnamefont {Z.}~\bibnamefont
  {Liu}}\ and\ \bibinfo {author} {\bibfnamefont {R.}~\bibnamefont
  {Slav{\'i}k}},\ }\bibfield  {title} {\enquote {\bibinfo {title} {Optical
  {{Injection Locking}}: {{From Principle}} to {{Applications}}},}\ }\href
  {\doibase 10.1109/JLT.2019.2945718} {\bibfield  {journal} {\bibinfo
  {journal} {Journal of Lightwave Technology}\ }\textbf {\bibinfo {volume}
  {38}},\ \bibinfo {pages} {43--59} (\bibinfo {year} {2020})}\BibitemShut
  {NoStop}%
\bibitem [{\citenamefont {Stover}\ and\ \citenamefont
  {Steier}(1966)}]{Stover1966}%
  \BibitemOpen
  \bibfield  {author} {\bibinfo {author} {\bibfnamefont {H.~L.}\ \bibnamefont
  {Stover}}\ and\ \bibinfo {author} {\bibfnamefont {W.~H.}\ \bibnamefont
  {Steier}},\ }\bibfield  {title} {\enquote {\bibinfo {title} {Locking of laser
  oscillators by light injection},}\ }\href {\doibase 10.1063/1.1754502}
  {\bibfield  {journal} {\bibinfo  {journal} {Applied Physics Letters}\
  }\textbf {\bibinfo {volume} {8}},\ \bibinfo {pages} {91--93} (\bibinfo {year}
  {1966})}\BibitemShut {NoStop}%
\bibitem [{\citenamefont {Kakarla}, \citenamefont {Schr{\"o}der},\ and\
  \citenamefont {Andrekson}(2018)}]{Kakarla2018}%
  \BibitemOpen
  \bibfield  {author} {\bibinfo {author} {\bibfnamefont {R.}~\bibnamefont
  {Kakarla}}, \bibinfo {author} {\bibfnamefont {J.}~\bibnamefont
  {Schr{\"o}der}}, \ and\ \bibinfo {author} {\bibfnamefont {P.~A.}\
  \bibnamefont {Andrekson}},\ }\bibfield  {title} {\enquote {\bibinfo {title}
  {Optical injection locking at sub nano-watt powers},}\ }\href {\doibase
  10.1364/OL.43.005769} {\bibfield  {journal} {\bibinfo  {journal} {Optics
  Letters}\ }\textbf {\bibinfo {volume} {43}},\ \bibinfo {pages} {5769--5772}
  (\bibinfo {year} {2018})}\BibitemShut {NoStop}%
\bibitem [{\citenamefont {Cai}, \citenamefont {Wake},\ and\ \citenamefont
  {Seeds}(1995)}]{Cai1995}%
  \BibitemOpen
  \bibfield  {author} {\bibinfo {author} {\bibfnamefont {B.}~\bibnamefont
  {Cai}}, \bibinfo {author} {\bibfnamefont {D.}~\bibnamefont {Wake}}, \ and\
  \bibinfo {author} {\bibfnamefont {A.}~\bibnamefont {Seeds}},\ }\bibfield
  {title} {\enquote {\bibinfo {title} {Microwave frequency synthesis using
  injection locked laser comb line selection},}\ }in\ \href {\doibase
  10.1109/LEOSST.1995.764171} {\emph {\bibinfo {booktitle} {{{IEEE}}/{{LEOS}}
  1995 {{Digest}} of the {{LEOS Summer Topical Meetings}}. {{Flat Panel Display
  Technology}}}}}\ (\bibinfo {year} {1995})\ pp.\ \bibinfo {pages}
  {13--14}\BibitemShut {NoStop}%
\bibitem [{\citenamefont {Moon}\ \emph {et~al.}(2006)\citenamefont {Moon},
  \citenamefont {Kim}, \citenamefont {Park},\ and\ \citenamefont
  {Park}}]{Moon2006}%
  \BibitemOpen
  \bibfield  {author} {\bibinfo {author} {\bibfnamefont {H.~S.}\ \bibnamefont
  {Moon}}, \bibinfo {author} {\bibfnamefont {E.~B.}\ \bibnamefont {Kim}},
  \bibinfo {author} {\bibfnamefont {S.~E.}\ \bibnamefont {Park}}, \ and\
  \bibinfo {author} {\bibfnamefont {C.~Y.}\ \bibnamefont {Park}},\ }\bibfield
  {title} {\enquote {\bibinfo {title} {Selection and amplification of modes of
  an optical frequency comb using a femtosecond laser injection-locking
  technique},}\ }\href {\doibase 10.1063/1.2374680} {\bibfield  {journal}
  {\bibinfo  {journal} {Applied Physics Letters}\ }\textbf {\bibinfo {volume}
  {89}},\ \bibinfo {pages} {181110} (\bibinfo {year} {2006})}\BibitemShut
  {NoStop}%
\bibitem [{\citenamefont {Ryu}\ \emph {et~al.}(2010)\citenamefont {Ryu},
  \citenamefont {Lee}, \citenamefont {Kim}, \citenamefont {Suh},\ and\
  \citenamefont {Moon}}]{Ryu2010}%
  \BibitemOpen
  \bibfield  {author} {\bibinfo {author} {\bibfnamefont {H.~Y.}\ \bibnamefont
  {Ryu}}, \bibinfo {author} {\bibfnamefont {S.~H.}\ \bibnamefont {Lee}},
  \bibinfo {author} {\bibfnamefont {E.~B.}\ \bibnamefont {Kim}}, \bibinfo
  {author} {\bibfnamefont {H.~S.}\ \bibnamefont {Suh}}, \ and\ \bibinfo
  {author} {\bibfnamefont {H.~S.}\ \bibnamefont {Moon}},\ }\bibfield  {title}
  {\enquote {\bibinfo {title} {A discretely tunable multifrequency source
  injection locked to a spectral-mode-filtered fiber laser comb},}\ }\href
  {\doibase 10.1063/1.3497080} {\bibfield  {journal} {\bibinfo  {journal}
  {Applied Physics Letters}\ }\textbf {\bibinfo {volume} {97}},\ \bibinfo
  {pages} {141107} (\bibinfo {year} {2010})}\BibitemShut {NoStop}%
\bibitem [{\citenamefont {Balakier}\ \emph {et~al.}(2014)\citenamefont
  {Balakier}, \citenamefont {Fice}, \citenamefont {van Dijk}, \citenamefont
  {Kervella}, \citenamefont {Carpintero}, \citenamefont {Seeds},\ and\
  \citenamefont {Renaud}}]{Balakier2014}%
  \BibitemOpen
  \bibfield  {author} {\bibinfo {author} {\bibfnamefont {K.}~\bibnamefont
  {Balakier}}, \bibinfo {author} {\bibfnamefont {M.~J.}\ \bibnamefont {Fice}},
  \bibinfo {author} {\bibfnamefont {F.}~\bibnamefont {van Dijk}}, \bibinfo
  {author} {\bibfnamefont {G.}~\bibnamefont {Kervella}}, \bibinfo {author}
  {\bibfnamefont {G.}~\bibnamefont {Carpintero}}, \bibinfo {author}
  {\bibfnamefont {A.~J.}\ \bibnamefont {Seeds}}, \ and\ \bibinfo {author}
  {\bibfnamefont {C.~C.}\ \bibnamefont {Renaud}},\ }\bibfield  {title}
  {\enquote {\bibinfo {title} {Optical injection locking of monolithically
  integrated photonic source for generation of high purity signals above 100
  {{GHz}}},}\ }\href {\doibase 10.1364/OE.22.029404} {\bibfield  {journal}
  {\bibinfo  {journal} {Optics Express}\ }\textbf {\bibinfo {volume} {22}},\
  \bibinfo {pages} {29404--29412} (\bibinfo {year} {2014})}\BibitemShut
  {NoStop}%
\bibitem [{\citenamefont {Shortiss}, \citenamefont {Shayesteh},\ and\
  \citenamefont {Peters}(2018)}]{Shortiss2018}%
  \BibitemOpen
  \bibfield  {author} {\bibinfo {author} {\bibfnamefont {K.~J.}\ \bibnamefont
  {Shortiss}}, \bibinfo {author} {\bibfnamefont {M.}~\bibnamefont {Shayesteh}},
  \ and\ \bibinfo {author} {\bibfnamefont {F.~H.}\ \bibnamefont {Peters}},\
  }\bibfield  {title} {\enquote {\bibinfo {title} {Modelling the effect of
  slave laser gain and frequency comb spacing on the selective amplification of
  injection locked semiconductor lasers},}\ }\href {\doibase
  10.1007/s11082-018-1317-3} {\bibfield  {journal} {\bibinfo  {journal}
  {Optical and Quantum Electronics}\ }\textbf {\bibinfo {volume} {50}},\
  \bibinfo {pages} {49} (\bibinfo {year} {2018})}\BibitemShut {NoStop}%
\bibitem [{\citenamefont {Kuse}\ \emph {et~al.}(2022)\citenamefont {Kuse},
  \citenamefont {Kuse}, \citenamefont {Minoshima},\ and\ \citenamefont
  {Minoshima}}]{Kuse2022}%
  \BibitemOpen
  \bibfield  {author} {\bibinfo {author} {\bibfnamefont {N.}~\bibnamefont
  {Kuse}}, \bibinfo {author} {\bibfnamefont {N.}~\bibnamefont {Kuse}}, \bibinfo
  {author} {\bibfnamefont {K.}~\bibnamefont {Minoshima}}, \ and\ \bibinfo
  {author} {\bibfnamefont {K.}~\bibnamefont {Minoshima}},\ }\bibfield  {title}
  {\enquote {\bibinfo {title} {Amplification and phase noise transfer of a
  {{Kerr}} microresonator soliton comb for low phase noise {{THz}} generation
  with a high signal-to-noise ratio},}\ }\href {\doibase 10.1364/OE.446903}
  {\bibfield  {journal} {\bibinfo  {journal} {Optics Express}\ }\textbf
  {\bibinfo {volume} {30}},\ \bibinfo {pages} {318--325} (\bibinfo {year}
  {2022})}\BibitemShut {NoStop}%
\bibitem [{\citenamefont {Gavrielides}(2014)}]{Gavrielides2014}%
  \BibitemOpen
  \bibfield  {author} {\bibinfo {author} {\bibfnamefont {A.}~\bibnamefont
  {Gavrielides}},\ }\bibfield  {title} {\enquote {\bibinfo {title} {Comb
  {{Injection}} and {{Sidebands Suppression}}},}\ }\href {\doibase
  10.1109/JQE.2014.2309532} {\bibfield  {journal} {\bibinfo  {journal} {IEEE
  Journal of Quantum Electronics}\ }\textbf {\bibinfo {volume} {50}},\ \bibinfo
  {pages} {364--371} (\bibinfo {year} {2014})}\BibitemShut {NoStop}%
\bibitem [{\citenamefont {Doumbia}\ \emph {et~al.}(2020)\citenamefont
  {Doumbia}, \citenamefont {Malica}, \citenamefont {Wolfersberger},
  \citenamefont {Panajotov},\ and\ \citenamefont {Sciamanna}}]{Doumbia2020}%
  \BibitemOpen
  \bibfield  {author} {\bibinfo {author} {\bibfnamefont {Y.}~\bibnamefont
  {Doumbia}}, \bibinfo {author} {\bibfnamefont {T.}~\bibnamefont {Malica}},
  \bibinfo {author} {\bibfnamefont {D.}~\bibnamefont {Wolfersberger}}, \bibinfo
  {author} {\bibfnamefont {K.}~\bibnamefont {Panajotov}}, \ and\ \bibinfo
  {author} {\bibfnamefont {M.}~\bibnamefont {Sciamanna}},\ }\bibfield  {title}
  {\enquote {\bibinfo {title} {Optical injection dynamics of frequency
  combs},}\ }\href {\doibase 10.1364/OL.381039} {\bibfield  {journal} {\bibinfo
   {journal} {Optics Letters}\ }\textbf {\bibinfo {volume} {45}},\ \bibinfo
  {pages} {435--438} (\bibinfo {year} {2020})}\BibitemShut {NoStop}%
\bibitem [{\citenamefont {{Albores-Mejia}}\ \emph {et~al.}(2015)\citenamefont
  {{Albores-Mejia}}, \citenamefont {Kaneko}, \citenamefont {Banno},
  \citenamefont {Uesaka}, \citenamefont {Shoji},\ and\ \citenamefont
  {Kuwatsuka}}]{Albores-Mejia2015}%
  \BibitemOpen
  \bibfield  {author} {\bibinfo {author} {\bibfnamefont {A.}~\bibnamefont
  {{Albores-Mejia}}}, \bibinfo {author} {\bibfnamefont {T.}~\bibnamefont
  {Kaneko}}, \bibinfo {author} {\bibfnamefont {E.}~\bibnamefont {Banno}},
  \bibinfo {author} {\bibfnamefont {K.}~\bibnamefont {Uesaka}}, \bibinfo
  {author} {\bibfnamefont {H.}~\bibnamefont {Shoji}}, \ and\ \bibinfo {author}
  {\bibfnamefont {H.}~\bibnamefont {Kuwatsuka}},\ }\bibfield  {title} {\enquote
  {\bibinfo {title} {Optical-{{Comb-Line Selection}} from a
  {{Low-Power}}/{{Low-OSNR Comb}} using a {{Low-Coherence Semiconductor Laser}}
  for {{Flexible Ultra-Dense Short Range Transceivers}}},}\ }in\ \href
  {\doibase 10.1364/OFC.2015.W2A.23} {\emph {\bibinfo {booktitle} {Optical
  {{Fiber Communication Conference}} (2015), Paper {{W2A}}.23}}}\ (\bibinfo
  {publisher} {{Optica Publishing Group}},\ \bibinfo {year} {2015})\ p.\
  \bibinfo {pages} {W2A.23}\BibitemShut {NoStop}%
\bibitem [{\citenamefont {Ishibashi}\ \emph {et~al.}(1997)\citenamefont
  {Ishibashi}, \citenamefont {Shimizu}, \citenamefont {Kodama}, \citenamefont
  {Ito}, \citenamefont {Nagatsuma},\ and\ \citenamefont
  {Furuta}}]{Ishibashi1997}%
  \BibitemOpen
  \bibfield  {author} {\bibinfo {author} {\bibfnamefont {T.}~\bibnamefont
  {Ishibashi}}, \bibinfo {author} {\bibfnamefont {N.}~\bibnamefont {Shimizu}},
  \bibinfo {author} {\bibfnamefont {S.}~\bibnamefont {Kodama}}, \bibinfo
  {author} {\bibfnamefont {H.}~\bibnamefont {Ito}}, \bibinfo {author}
  {\bibfnamefont {T.}~\bibnamefont {Nagatsuma}}, \ and\ \bibinfo {author}
  {\bibfnamefont {T.}~\bibnamefont {Furuta}},\ }\bibfield  {title} {\enquote
  {\bibinfo {title} {Uni-{{Traveling-Carrier Photodiodes}}},}\ }in\ \href
  {\doibase 10.1364/UEO.1997.UC3} {\emph {\bibinfo {booktitle} {Ultrafast
  {{Electronics}} and {{Optoelectronics}} (1997), Paper {{UC3}}}}}\ (\bibinfo
  {publisher} {{Optica Publishing Group}},\ \bibinfo {year} {1997})\ p.\
  \bibinfo {pages} {UC3}\BibitemShut {NoStop}%
\bibitem [{\citenamefont {Ishibashi}\ and\ \citenamefont
  {Ito}(2020)}]{Ishibashi2020}%
  \BibitemOpen
  \bibfield  {author} {\bibinfo {author} {\bibfnamefont {T.}~\bibnamefont
  {Ishibashi}}\ and\ \bibinfo {author} {\bibfnamefont {H.}~\bibnamefont
  {Ito}},\ }\bibfield  {title} {\enquote {\bibinfo {title}
  {Uni-traveling-carrier photodiodes},}\ }\href {\doibase 10.1063/1.5128444}
  {\bibfield  {journal} {\bibinfo  {journal} {Journal of Applied Physics}\
  }\textbf {\bibinfo {volume} {127}},\ \bibinfo {pages} {031101} (\bibinfo
  {year} {2020})}\BibitemShut {NoStop}%
\bibitem [{\citenamefont {Kippenberg}\ \emph {et~al.}(2018)\citenamefont
  {Kippenberg}, \citenamefont {Gaeta}, \citenamefont {Lipson},\ and\
  \citenamefont {Gorodetsky}}]{Kippenberg2018}%
  \BibitemOpen
  \bibfield  {author} {\bibinfo {author} {\bibfnamefont {T.~J.}\ \bibnamefont
  {Kippenberg}}, \bibinfo {author} {\bibfnamefont {A.~L.}\ \bibnamefont
  {Gaeta}}, \bibinfo {author} {\bibfnamefont {M.}~\bibnamefont {Lipson}}, \
  and\ \bibinfo {author} {\bibfnamefont {M.~L.}\ \bibnamefont {Gorodetsky}},\
  }\bibfield  {title} {\enquote {\bibinfo {title} {Dissipative {{Kerr}}
  solitons in optical microresonators},}\ }\href {\doibase
  10.1126/science.aan8083} {\bibfield  {journal} {\bibinfo  {journal}
  {Science}\ }\textbf {\bibinfo {volume} {361}},\ \bibinfo {pages} {eaan8083}
  (\bibinfo {year} {2018})}\BibitemShut {NoStop}%
\bibitem [{\citenamefont {Shen}\ \emph {et~al.}(2020)\citenamefont {Shen},
  \citenamefont {Chang}, \citenamefont {Liu}, \citenamefont {Wang},
  \citenamefont {Yang}, \citenamefont {Xiang}, \citenamefont {Wang},
  \citenamefont {He}, \citenamefont {Liu}, \citenamefont {Xie}, \citenamefont
  {Guo}, \citenamefont {Kinghorn}, \citenamefont {Wu}, \citenamefont {Ji},
  \citenamefont {Kippenberg}, \citenamefont {Vahala},\ and\ \citenamefont
  {Bowers}}]{Shen2020}%
  \BibitemOpen
  \bibfield  {author} {\bibinfo {author} {\bibfnamefont {B.}~\bibnamefont
  {Shen}}, \bibinfo {author} {\bibfnamefont {L.}~\bibnamefont {Chang}},
  \bibinfo {author} {\bibfnamefont {J.}~\bibnamefont {Liu}}, \bibinfo {author}
  {\bibfnamefont {H.}~\bibnamefont {Wang}}, \bibinfo {author} {\bibfnamefont
  {Q.-F.}\ \bibnamefont {Yang}}, \bibinfo {author} {\bibfnamefont
  {C.}~\bibnamefont {Xiang}}, \bibinfo {author} {\bibfnamefont {R.~N.}\
  \bibnamefont {Wang}}, \bibinfo {author} {\bibfnamefont {J.}~\bibnamefont
  {He}}, \bibinfo {author} {\bibfnamefont {T.}~\bibnamefont {Liu}}, \bibinfo
  {author} {\bibfnamefont {W.}~\bibnamefont {Xie}}, \bibinfo {author}
  {\bibfnamefont {J.}~\bibnamefont {Guo}}, \bibinfo {author} {\bibfnamefont
  {D.}~\bibnamefont {Kinghorn}}, \bibinfo {author} {\bibfnamefont
  {L.}~\bibnamefont {Wu}}, \bibinfo {author} {\bibfnamefont {Q.-X.}\
  \bibnamefont {Ji}}, \bibinfo {author} {\bibfnamefont {T.~J.}\ \bibnamefont
  {Kippenberg}}, \bibinfo {author} {\bibfnamefont {K.}~\bibnamefont {Vahala}},
  \ and\ \bibinfo {author} {\bibfnamefont {J.~E.}\ \bibnamefont {Bowers}},\
  }\bibfield  {title} {\enquote {\bibinfo {title} {Integrated turnkey soliton
  microcombs},}\ }\href {\doibase 10.1038/s41586-020-2358-x} {\bibfield
  {journal} {\bibinfo  {journal} {Nature}\ }\textbf {\bibinfo {volume} {582}},\
  \bibinfo {pages} {365--369} (\bibinfo {year} {2020})}\BibitemShut {NoStop}%
\bibitem [{\citenamefont {Billah}\ \emph {et~al.}(2018)\citenamefont {Billah},
  \citenamefont {Blaicher}, \citenamefont {Hoose}, \citenamefont {Dietrich},
  \citenamefont {{Marin-Palomo}}, \citenamefont {Lindenmann}, \citenamefont
  {Nesic}, \citenamefont {Hofmann}, \citenamefont {Troppenz}, \citenamefont
  {Moehrle}, \citenamefont {Randel}, \citenamefont {Freude},\ and\
  \citenamefont {Koos}}]{Billah2018}%
  \BibitemOpen
  \bibfield  {author} {\bibinfo {author} {\bibfnamefont {M.~R.}\ \bibnamefont
  {Billah}}, \bibinfo {author} {\bibfnamefont {M.}~\bibnamefont {Blaicher}},
  \bibinfo {author} {\bibfnamefont {T.}~\bibnamefont {Hoose}}, \bibinfo
  {author} {\bibfnamefont {P.-I.}\ \bibnamefont {Dietrich}}, \bibinfo {author}
  {\bibfnamefont {P.}~\bibnamefont {{Marin-Palomo}}}, \bibinfo {author}
  {\bibfnamefont {N.}~\bibnamefont {Lindenmann}}, \bibinfo {author}
  {\bibfnamefont {A.}~\bibnamefont {Nesic}}, \bibinfo {author} {\bibfnamefont
  {A.}~\bibnamefont {Hofmann}}, \bibinfo {author} {\bibfnamefont
  {U.}~\bibnamefont {Troppenz}}, \bibinfo {author} {\bibfnamefont
  {M.}~\bibnamefont {Moehrle}}, \bibinfo {author} {\bibfnamefont
  {S.}~\bibnamefont {Randel}}, \bibinfo {author} {\bibfnamefont
  {W.}~\bibnamefont {Freude}}, \ and\ \bibinfo {author} {\bibfnamefont
  {C.}~\bibnamefont {Koos}},\ }\bibfield  {title} {\enquote {\bibinfo {title}
  {Hybrid integration of silicon photonics circuits and {{InP}} lasers by
  photonic wire bonding},}\ }\href {\doibase 10.1364/OPTICA.5.000876}
  {\bibfield  {journal} {\bibinfo  {journal} {Optica}\ }\textbf {\bibinfo
  {volume} {5}},\ \bibinfo {pages} {876--883} (\bibinfo {year}
  {2018})}\BibitemShut {NoStop}%
\bibitem [{\citenamefont {Maes}\ \emph {et~al.}(2023)\citenamefont {Maes},
  \citenamefont {Lemey}, \citenamefont {Roelkens}, \citenamefont {Zaknoune},
  \citenamefont {Avramovic}, \citenamefont {Okada}, \citenamefont
  {Szriftgiser}, \citenamefont {Peytavit}, \citenamefont {Ducournau},\ and\
  \citenamefont {Kuyken}}]{Maes2023}%
  \BibitemOpen
  \bibfield  {author} {\bibinfo {author} {\bibfnamefont {D.}~\bibnamefont
  {Maes}}, \bibinfo {author} {\bibfnamefont {S.}~\bibnamefont {Lemey}},
  \bibinfo {author} {\bibfnamefont {G.}~\bibnamefont {Roelkens}}, \bibinfo
  {author} {\bibfnamefont {M.}~\bibnamefont {Zaknoune}}, \bibinfo {author}
  {\bibfnamefont {V.}~\bibnamefont {Avramovic}}, \bibinfo {author}
  {\bibfnamefont {E.}~\bibnamefont {Okada}}, \bibinfo {author} {\bibfnamefont
  {P.}~\bibnamefont {Szriftgiser}}, \bibinfo {author} {\bibfnamefont
  {E.}~\bibnamefont {Peytavit}}, \bibinfo {author} {\bibfnamefont
  {G.}~\bibnamefont {Ducournau}}, \ and\ \bibinfo {author} {\bibfnamefont
  {B.}~\bibnamefont {Kuyken}},\ }\bibfield  {title} {\enquote {\bibinfo {title}
  {High-speed uni-traveling-carrier photodiodes on silicon nitride},}\ }\href
  {\doibase 10.1063/5.0119244} {\bibfield  {journal} {\bibinfo  {journal} {APL
  Photonics}\ }\textbf {\bibinfo {volume} {8}},\ \bibinfo {pages} {016104}
  (\bibinfo {year} {2023})}\BibitemShut {NoStop}%
\bibitem [{\citenamefont {Tetsumoto}\ and\ \citenamefont
  {Rolland}(2022)}]{Tetsumoto2022}%
  \BibitemOpen
  \bibfield  {author} {\bibinfo {author} {\bibfnamefont {T.}~\bibnamefont
  {Tetsumoto}}\ and\ \bibinfo {author} {\bibfnamefont {A.}~\bibnamefont
  {Rolland}},\ }\href {\doibase 10.48550/arXiv.2210.15881} {\enquote {\bibinfo
  {title} {300 {{GHz}} wireless link based on an integrated {{Kerr}} soliton
  comb},}\ } (\bibinfo {year} {2022}),\ \Eprint
  {http://arxiv.org/abs/2210.15881} {arXiv:2210.15881 [physics]} \BibitemShut
  {NoStop}%
\bibitem [{\citenamefont {Stone}\ \emph {et~al.}(2018)\citenamefont {Stone},
  \citenamefont {Briles}, \citenamefont {Drake}, \citenamefont {Spencer},
  \citenamefont {Carlson}, \citenamefont {Diddams},\ and\ \citenamefont
  {Papp}}]{Stone2018}%
  \BibitemOpen
  \bibfield  {author} {\bibinfo {author} {\bibfnamefont {J.~R.}\ \bibnamefont
  {Stone}}, \bibinfo {author} {\bibfnamefont {T.~C.}\ \bibnamefont {Briles}},
  \bibinfo {author} {\bibfnamefont {T.~E.}\ \bibnamefont {Drake}}, \bibinfo
  {author} {\bibfnamefont {D.~T.}\ \bibnamefont {Spencer}}, \bibinfo {author}
  {\bibfnamefont {D.~R.}\ \bibnamefont {Carlson}}, \bibinfo {author}
  {\bibfnamefont {S.~A.}\ \bibnamefont {Diddams}}, \ and\ \bibinfo {author}
  {\bibfnamefont {S.~B.}\ \bibnamefont {Papp}},\ }\bibfield  {title} {\enquote
  {\bibinfo {title} {Thermal and {{Nonlinear Dissipative-Soliton Dynamics}} in
  {{Kerr-Microresonator Frequency Combs}}},}\ }\href {\doibase
  10.1103/PhysRevLett.121.063902} {\bibfield  {journal} {\bibinfo  {journal}
  {Physical Review Letters}\ }\textbf {\bibinfo {volume} {121}},\ \bibinfo
  {pages} {063902} (\bibinfo {year} {2018})}\BibitemShut {NoStop}%
\bibitem [{\citenamefont {Metcalf}\ \emph {et~al.}(2013)\citenamefont
  {Metcalf}, \citenamefont {{Torres-Company}}, \citenamefont {Leaird},\ and\
  \citenamefont {Weiner}}]{Metcalf2013}%
  \BibitemOpen
  \bibfield  {author} {\bibinfo {author} {\bibfnamefont {A.~J.}\ \bibnamefont
  {Metcalf}}, \bibinfo {author} {\bibfnamefont {V.}~\bibnamefont
  {{Torres-Company}}}, \bibinfo {author} {\bibfnamefont {D.~E.}\ \bibnamefont
  {Leaird}}, \ and\ \bibinfo {author} {\bibfnamefont {A.~M.}\ \bibnamefont
  {Weiner}},\ }\bibfield  {title} {\enquote {\bibinfo {title} {High-{{Power
  Broadly Tunable Electrooptic Frequency Comb Generator}}},}\ }\href {\doibase
  10.1109/JSTQE.2013.2268384} {\bibfield  {journal} {\bibinfo  {journal} {IEEE
  Journal of Selected Topics in Quantum Electronics}\ }\textbf {\bibinfo
  {volume} {19}},\ \bibinfo {pages} {231--236} (\bibinfo {year}
  {2013})}\BibitemShut {NoStop}%
\bibitem [{201(2017)}]{2017}%
  \BibitemOpen
  \bibfield  {title} {\enquote {\bibinfo {title} {{{IEEE Standard}} for {{High
  Data Rate Wireless Multi-Media Networks}}\textendash{{Amendment}} 2: 100
  {{Gb}}/s {{Wireless Switched Point-to-Point Physical Layer}}},}\ }\href
  {\doibase 10.1109/IEEESTD.2017.8066476} {\bibfield  {journal} {\bibinfo
  {journal} {IEEE Std 802.15.3d-2017 (Amendment to IEEE Std 802.15.3-2016 as
  amended by IEEE Std 802.15.3e-2017)}\ ,\ \bibinfo {pages} {1--55}} (\bibinfo
  {year} {2017})}\BibitemShut {NoStop}%
\bibitem [{\citenamefont {Petrov}, \citenamefont {Kurner},\ and\ \citenamefont
  {Hosako}(2020)}]{Petrov2020}%
  \BibitemOpen
  \bibfield  {author} {\bibinfo {author} {\bibfnamefont {V.}~\bibnamefont
  {Petrov}}, \bibinfo {author} {\bibfnamefont {T.}~\bibnamefont {Kurner}}, \
  and\ \bibinfo {author} {\bibfnamefont {I.}~\bibnamefont {Hosako}},\
  }\bibfield  {title} {\enquote {\bibinfo {title} {{{IEEE}} 802.15.3d: {{First
  Standardization Efforts}} for {{Sub-Terahertz Band Communications}} toward
  {{6G}}},}\ }\href {\doibase 10.1109/MCOM.001.2000273} {\bibfield  {journal}
  {\bibinfo  {journal} {IEEE Communications Magazine}\ }\textbf {\bibinfo
  {volume} {58}},\ \bibinfo {pages} {28--33} (\bibinfo {year}
  {2020})}\BibitemShut {NoStop}%
\end{thebibliography}%

\end{document}